\documentclass[11pt,a4paper]{article}
\pdfoutput=1
\usepackage{jheppub}
\usepackage{caption}
\usepackage{subcaption}

\newcommand{\fig}[1]{Fig.\ \ref{#1}}
\newcommand{\tab}[1]{Table\ \ref{#1}}

\newcommand{\eqq}[1]{Eq.~(\ref{#1})}

\newcommand{\be}{\begin{equation}}

\newcommand{\ee}{\end{equation}}

\newcommand{\bea}{\begin{eqnarray}}
\newcommand{\eea}{\end{eqnarray}}

\def\phiH{{\phi_H}}
\def\phiM{\phi_M}

\def\phit{\tilde \phi}
 
\begin{document}
\begin{titlepage}
\thispagestyle{empty}

\vspace{2cm}
\begin{center}
\font\titlerm=cmr10 scaled\magstep4
\font\titlei=cmmi10
scaled\magstep4 \font\titleis=cmmi7 scaled\magstep4 {\Large{\textbf{Jet suppression in non-conformal plasma using AdS/CFT}
\\}}
\setcounter{footnote}{0}
\vspace{1.5cm} \noindent{{
 Sara Heshmatian$^{a}$\footnote{e-mail:heshmatian@bzte.ac.ir }, Razieh Morad$^{b,c}$\footnote{e-mail:rmorad@tlabs.ac.za}, Mahmood Akbari$^{b,c}$\footnote{e-mail:makbari@tlabs.ac.za}
}}\\
\vspace{0.8cm}

{\it $^{a}$ Department of Engineering Sciences and Physics, Buein Zahra Technical University, Buein Zahra, Qazvin, Iran\\}
{\it $^{b}$ UNESCO-UNISA Africa Chair in Nanosciences/Nanotechnology, College of Graduate Studies, University of South Africa (UNISA), Muckleneuk Ridge, P.O.Box 392, Pretoria, South Africa \\}
{\it $^{c}$ Nanosciences African Network, Materials Research Department, iThemba LABS, P.O.Box 722, National Research Foundation, South Africa}

\vspace*{.4cm}
\end{center}
\vskip 2em
\setcounter{footnote}{0}
\begin{abstract}

In this paper, we study suppression of light quark in strongly coupled non-conformal plasmas using the AdS/CFT correspondence. The well-known falling string profile in the bulk is considered as light quark moving through the plasma. The maximum distance which string with energy E can travel before falling through the horizon is interpreted as thermalization distance of light quark in the hot-strongly coupled plasma. Our numerical results show that the thermalization distance of light quark increases by increasing deviation from conformal invariance. The relation between this distance and the energy of quark and the temperature of the plasma is analyzed numerically. The jet quenching parameter is also calculated in this non-conformal background and it is found that the jet quenching parameter is decreased by increasing the non-conformality. Our results are compared with the results of  $\mathcal N=4$ SYM theory and also some available experimental data.

\end{abstract}
\end{titlepage}
\tableofcontents

\section{Introduction}

Recent experiments performed at the Relativistic Heavy Ion Collider (RHIC) and the Large Hadron Collider (LHC) has provided spectacular evidence that suggests that a deconfined state of hadronic matter has been formed which is called quark-gluon plasma (QGP) \cite{Adams:2005dq,Adcox:2004mh,Arsene:2004fa,Back:2004je}. The experimental discovery, e.g, a very small ratio of shear viscosity to entropy density \cite{Policastro:2001yc,Kovtun:2004de}, quenching of high energy partons with large transverse momentum, and elliptic flow etc., indicates that QGP is a strongly coupled plasma whose dynamics after collision is dominated by non-perturbative effects \cite{Baier:1996kr,Eskola:2004cr}. The perturbative QCD works only in the weak coupling regime, while lattice QCD is the proper tool for understanding the static equilibrium thermodynamics of such strongly coupled plasma, it does not allow us to compute some dynamical quantities like transport coefficients, drag force, and jet quenching parameter etc., in the strong coupling regime. 

Recently, a novel tool called the ``AdS/CFT correspondence" \cite{Maldacena:1997re,Witten:1998qj,Gubser:1998bc,Aharony:1999ti,CasalderreySolana:2011us} provide a powerful tool to study the strongly coupled plasma. The most studied example in the context of AdS/CFT correspondence \cite{Shuryak:2008eq,Shuryak:2004cy,Baier:1996kr} is the duality between the $\mathcal{N} = 4\, SU(N_c)$ super-Yang-Mills theory and type IIB string theory on $AdS_5 \times S^5$, which allows one to study this strongly coupled gauge theory in the large $N_c$ limit and large ’t Hooft coupling $\lambda = g_{Y M}^2\,N_c$ .The ratio of shear viscosity over entropy density obtained from this duality is small $\eta/s = 1/4\pi$ \cite{Policastro:2001yc,Kovtun:2003wp,Buchel:2003tz} which is consistent with the experimental data \cite{Teaney:2003kp}.  Also, the strong collective behavior observed in very small systems, such as Au-Au collisions at RHIC \cite{Ackermann:2000tr,Adler:2003kt,Back:2004mh}, Pb-Pb \cite{ATLAS:2012at,Chatrchyan:2012ta,Aamodt:2010pa,Adam:2016izf}, p-Pb \cite{Aad:2014lta,Khachatryan:2015waa,Abelev:2014mda} and p-p \cite{Aad:2015gqa} collisions at the LHC are well described by hydrodynamics from the time and distance scales as a fraction of the (local) inverse temperature \cite{Heller:2011ju,Chesler:2009cy,Chesler:2015wra,Chesler:2013lia,Casalderrey-Solana:2013sxa,Casalderrey-Solana:2013aba} which is consistent with holographic results \cite{Chesler:2016ceu,Chesler:2015bba}.

Despite the fact that the initial temperatures in the most central, high-energy collisions at  in the range of quasi-conformal regime the non-conformal effects become important in the subsequent evolution and cooling of the QGP after production. Also, the initial temperature is smaller in the off-central or lower-energy collisions both at the LHC and RHIC. Lattice data indicate that in the temperature range available at high-energy heavy ion collisions (temperatures a few times larger than the critical temperature), the QGP is not a fully conformal fluid and the bulk viscosity (a purely non-conformal effect) is needed for highly-precise extraction of the shear viscosity of the QGP \cite{Ryu:2015vwa}. Furthermore, hydrodynamics including non-conformal effects successfully described the smaller system such as p-Pb \cite{Bozek:2011if} and p-p \cite{Schenke:2014zha,Habich:2015rtj} collisions \cite{Jeon:2015dfa}.

There are many attempts to extend the original duality to the theories close to QCD or QGP either uses a top-down \cite{Polchinski:2000uf,Karch:2002sh,Sakai:2004cn} or a bottom-up \cite{Gursoy:2007cb,Galow:2009kw} construction. In the latter, the equations of motion of a five-dimensional supergravity action coupled to a matter content of fields will be solved. One may break the conformal invariance even at zero temperature by coupling a scalar field at pure gravity in AdS \cite{Attems:2016ugt} which is dual to a CFT deformed by a source $\Lambda$ for a dimension-three operator. This source breaks the scale invariance explicitly and triggers a non-trivial Renormalization Group (RG) flow from ultraviolet (UV) fixed point to an infrared (IR) fixed point. Usually, the scalar potential coupled to the gravity is chosen to mimic the detailed properties of QCD,  e.g. \cite{Gubser:2008ny}. However, the authors in \cite{Attems:2016ugt} chose their potential by simplicity. The UV fixed point is needed to guarantee that we are in the regime that the holographic duality is best understood and the bulk metric is asymptotically AdS. The IR fixed point guarantees that the solutions are regular in the interior and the zero-temperature solution is smooth in the deep IR. Although turning on a source for a relevant operator is the simplest way to break conformal invariance, this simple model exhibits a rich phenomenology. For example, the relaxation of small-amplitude and homogeneous perturbations are studied by computing the spectrum of quasi-normal modes (QNM) and it is shown that the dominant channel for relaxation in this approximation depends on the value of the ratio $T/\Lambda$, with T the temperature of the system.  At small $T/\Lambda$ the system first EoSizes (the evolution of the energy density and the average pressure towards asymptotic values related to one another by the equation of state (EoS)) and then isotropises. While at large $T/\Lambda$, the system first isotropises and subsequently EoSizes  \cite{Attems:2016ugt}.

One of the interesting properties of QGP is jet quenching processes in which a high energy parton propagates through the medium, interact with medium and consequently lose energy.  Analyses the energy loss of these energetic partons as they travel throw QGP may reveal extremely valuable information about the dynamics of the plasma and exhibit distinctive properties such as jet-quenching which can clearly be observed at RHIC \cite{Adams:2005dq,Adcox:2004mh,Arsene:2004fa,Back:2004je} and more recently LHC \cite{Yin:2013zea,Aad:2010bu,Chatrchyan:2011sx}. 

In the AdS/CFT correspondence, an external heavy quark with infinite mass can be introduced by adding a fundamental string attached to the flavor brane on the AdS boundary. The endpoint of the string indicates the heavy quark while the string itself can be considered as a gluonic cloud around the quark. The mass of quark is proportional to the length of the string, such that light quark or massless quark is mapped into a string attached to a flavor brane which is extended from the boundary to the horizon \cite{Herzog:2006gh,Gubser:2006bz,Horowitz:2009pw,Fadafan:2008bq,Horowitz:2015dta}. 

A pair of light quark-anti quark is modelled by an initially point-like open string created close to the boundary with endpoints that are free to fly apart \cite{Chesler:2008uy,Morad:2014xla}. The initial conditions of the string imply that the string extends in a direction parallel to the boundary as it falls toward the black hole horizon, so it is called the falling string. The maximum stopping distance of the falling string (the maximum distance which a quark with initial energy E can travel) can be used as a phenomenological guideline to estimate the stopping power of the strongly- coupled plasma. This quantity is calculated numerically for many different sets of string initial conditions in a thermal N = 4 SYM plasma at a temperature T and it is found that the maximum penetration depth scales as $x_{max}=\frac{C}{T}(\frac{E}{T\sqrt{\lambda}})^{n_{eff}}$ with $n_{eff}=1/3$ \cite{Chesler:2008uy}. The analogous computation is done in a plasma with non-conformal effect \cite{Ficnar:2011yj} or with anisotropy effect \cite{BitaghsirFadafan:2017tci} and it is shown that the power of $n_{eff}$ deviates from its conformal isotropic cousin. The constant of proportionality is also important for phenomenological applications as it determines the overall strength of jet quenching. Here, we analyze the maximum stopping distance of light quark by numerical computation of stopping distance for many different sets of string initial conditions in the non-conformal background introduced in \cite{Attems:2016ugt}.

Another interesting experimental observables associated with quark energy lost in the dense hot QGP is the transport coefficient $\hat{q}$ which is called ``jet quenching". This observable is defined as the average transverse momentum square transferred from the traversing parton, per unit mean free path \cite{DEramo:2010wup}. As mentioned before, the quark--antiquark pair in the context of AdS/CFT is mapped to a fundamental string with both endpoints attach on the AdS boundary. The string hangs down to the bulk along the radial direction. The jet quenching parameter $\hat{q}$ is related to the thermal expectation value of the light-like Wilson loop operator \cite{Liu:2006he} which correspond to the trajectory of two endpoints of the string. Several attempts were made to compute this parameter using the AdS/CFT correspondence \cite{Liu:2006he,Caceres:2006as,Buchel:2006bv,VazquezPoritz:2006ba,Nakano:2006js,Avramis:2006ip,Gao:2006uf,Armesto:2006zv,Lin:2006au,Sadeghi:2013dga,Gursoy:2010fj,Cai:2012eh,Wang:2016noh,DeWolfe:2009vs,Fadafan:2008uv,Horowitz:2017nbm}. In this paper, we also study the jet quenching parameter in the introduced non-conformal background. Our results are in good agreement with experimental expectation. 

This paper is organised as follows: In section \ref{section:nonconBG} we briefly review the non-conformal background introduced in \cite{Attems:2016ugt}.  In section \ref{section:LightJet} we discuss the generic falling string solutions and calculate the maximum stopping distance of light quark in the strongly coupled non-conformal plasma. In section \ref{section:JQ}, we calculate the jet quenching parameter and section \ref{section:Discussion} is devoted to conclusion and discussion.


\section{Non-conformal holographic model}
\label{section:nonconBG}

In this section we review the non-conformal holographic model presented in \cite{Attems:2016ugt}. The action for the five-dimensional Einstein gravity coupled to a scalar field is
\begin{equation}
\label{eq:action}
S=\frac{2}{\kappa_5^2} \int d^5 x \sqrt{-g} \left[ \frac{1}{4} \mathcal{R}  - \frac{1}{2} \left( \nabla \phi \right) ^2 - V(\phi) \right ] \, ,
\end{equation}
where $\kappa_5$ is is the five-dimensional Newton constant. The potential considered in this model has the following non-trivial rather simple form  
\begin{equation}
\label{eq:pot}
L^2 V=-3 -\frac{3}{2} \phi^2 - \frac{1}{3} \phi^4 + \left( \frac{1}{3 \phi_M^2} +  \frac{1}{2 \phi_M^4}\right) \phi^6-\frac{1}{12 \phi_M^4} \phi^8 \, .
\end{equation}
This potential possess a maximum at $\phi=0$ (UV fixed point)  and a minimum at \mbox{$\phi=\phi_M$} (IR fixed point) corresponding to two AdS solutions. $L$ is the radius of the corresponding AdS solution at the UV whereas the radius of the IR AdS is 
\begin{equation}
\label{eq:LIR}
L_{\rm IR} = \sqrt{- \frac{3}{V\left(\phi_M\right)}} = \frac{1}{1+ \frac{1}{6} \phi_M^2} L \, .
\end{equation}
The number of degrees of freedom decreases along the flow (from a UV to an IR) because of the fact that  $L_{\rm IR} < L$.

The vacuum solutions to the Einstein equations can be easily found 
\be
\label{eq:adsvac}
ds^2 = e^{2 A(r)} \left(-d t^2 + d{\bf x}^2\right) + dr^2 \, ,
\ee
where 
\begin{eqnarray}
\label{eq:vacmetricsol}
e^{2 A}&=& \frac{\Lambda^2 L^2}{\phi^2} \,
  \left(1- \frac{\phi ^2}{\phi _M^2} \right)^{\frac{\phi_M^2}{6}+1} \, 
  e^{-\frac{\phi ^2}{6}}  \,,
\\[2mm]
\label{eq:phisol}
\phi(r)&=& \frac{\Lambda L \, e^{-r/L}}{\sqrt{1+ \frac{\Lambda^2 L^2}{\phi_M^2}e^{-2 r/L} }} \,.
\end{eqnarray}
Here, $\Lambda$ is an arbitrary constant that breaks conformal invariance. 
The scalar field is dual to a scalar operator in the dual gauge theory $\O$, with different dimension at the UV and IR fixed points. The dimension of this operator at the UV fixed point is $\Delta_{UV} = 3$ while the operator $\O$ at the IR fixed point has dimension 
\begin{equation}
\label{IRdim}
\Delta_\textrm{IR}=
6\, \left( 1+\frac{\phiM^2}{9}\right) \left(1+\frac{\phiM^2}{6} \right)^{-1}\,.
\end{equation}
So, by increasing $\phiM$ the dimension of the scalar operator at the IR fixed point decreases and finally reaches $\Delta_{IR}=4$ at $\phiM \rightarrow \infty$. 
One can determine the vacuum expectation values (VEV) of the stress tensor and the scalar operator by studying the behavior of the metric and the scalar field near the boundary. Using the new variable $u$ as $u=L\, e^{-r/L}$, the metric can be rewritten as  \cite{Bianchi:2001kw}
\be
ds^2 = \frac{L^2}{u^2} \left (d u^2 + g_{\mu \nu} \, dx^\mu dx^\nu\right) \, .
\ee
The expectation values of the field theory operators can be readily determined by expanding the metric and the scalar field in powers of $u$ in the $u \rightarrow 0$ limit as follows \cite{Attems:2016ugt}
\bea
\label{eq:Texp}
 \left< T_{\mu \nu} \right> &=& 
 \frac{2 L^3}{\kappa_5^2} 
 \left[ g^{(4)}_{\mu \nu}  + \left(\Lambda^2 \,\phi_2 -  \frac{\Lambda^4}{18} + \frac{\Lambda^4}{4\phiM^2}\right) \eta_{\mu \nu}\right] \, , \\[2mm]
\label{eq:Oexp}
\left< \O \right> &=& - \frac{2 L^3}{\kappa_5^2}\,  \left(2 \Lambda \phi_2  + \frac{\Lambda^3}{\phiM^2} \right)\,,
\eea
which imply the Ward identity for the trace of the stress tensor
\be
\label{eq:TTrace0}
\left<T^{\mu}_\mu\right>= - \Lambda \left< \O \right> \, .
\ee
Since the scalar field is a monotonic function of $r$, we can use the $\phi$ as a coordinate and consider the following ansatz for black brane solutions of the action in the Eddington-Finkelstein form
\be
\label{eq:blackads}
ds^2= e^{2 A}\left(-h(\phi) d\tilde t\, ^ 2 + d{\bf x}^2 \right) -2 e^{A+B} L \, d\tilde t \,d\phi \, ,
\ee
where $h(\phi)$ is the blackening factor which is zero at $\phi=\phi_H$. The boundary of space is at $\phi=0$ and $\phi_H$ is the location of the horizon.  
With this ansatz, the equations of motion can be written as   
 \begin{subequations}
 \label{eq:Einstein}
 \begin{alignat}{4}
 \label{eq:tt}
 A''(\phi )-A'(\phi ) B'(\phi )+\frac{2}{3}&=&0  \,,  \\
 \label{eq:xx}
 4 A'(\phi ) h'(\phi )-B'(\phi ) h'(\phi )+h''(\phi
   )&=&0 \,, \\
   \label{eq:phiphi}
 \frac{3}{2} A'(\phi ) h'(\phi )+h(\phi ) \left(6
   A'(\phi )^2-1\right)+2 e^{2 B(\phi )}
 L^2 V(\phi )&=&0  \,,\\
 \label{eq:VV}
 4 A'(\phi )-B'(\phi )-\frac{e^{2 B(\phi )} L^2
  V'(\phi )}{h(\phi )}+\frac{h'(\phi
   )}{h(\phi )}&=&0 \, ,
 \end{alignat}
  \end{subequations}
where primes denote $d/d\phi$. It is shown that  these four equations can be combined into a master equation of the form \cite{Attems:2016ugt}
\bea
\label{eq:master}
\frac{G'(\phi )}{G(\phi )+\frac{4 V(\phi )}{3 V'(\phi )}}
=
\frac{d}{d\phi} \log \left(
\frac{1}{3 G(\phi )}-2 G(\phi )+\frac{G'(\phi )}{2 G(\phi )}-\frac{G'(\phi )}{2 \left(G(\phi )+\frac{4 V(\phi )}{3 V'(\phi )}\right)} 
\right) \, ,
\eea
where $G(\phi)= \frac{d}{d\phi} A(\phi)$ is assumed to be a smooth generating function which can be integrated to obtain $A(\phi)$ as
\be
\label{eq:A}
A(\phi)=A_0 + \int_{\phi_0}^{\phi} d\phit\, G(\phit) \, .
\ee
Integrating the first three equations of Eq. \eqq{eq:Einstein} leads to 
 \begin{subequations}
  \begin{alignat}{4}
\label{eq:B}
B(\phi)=B_0  + \int_{\phi_0}^{\phi} d\phit\, \frac{G'(\phit)+2/3}{G(\phit)} \, ,&&\\
\label{eq:h}
h(\phi)=h_0  +\,h_1\, \int_{\phi_0}^{\phi}  \, d\phit\, e^{-4\, A(\phit)+B(\phit)}\, ,&& \\
\label{eq:V}
V(\phi) =  {h(\phi) \, e^{-2B(\phi)} \over {2 L^2}} \left( 1 -
    6 G(\phi)^2 -  {3\, G(\phi)\,h'(\phi) \over{2\,h(\phi)}} \right)\, .
\end{alignat}
\end{subequations}
Using the fact that $h$ has a simple zero at the horizon, the \eqq{eq:phiphi} and \eqq{eq:VV} give rise to 
\begin{subequations}
  \begin{alignat}{4}
\label{eq:VH}
V(\phi_H)=-\frac{3}{4\, L^2}  G(\phiH) h'(\phiH)\,e^{-2B(\phiH)} \,, &&\\
\label{eq:VpH}
V'(\phi_H)=\frac{1}{ L^2}   h'(\phiH)\, e^{-2B(\phiH)} \, ,&& 
\end{alignat}
\end{subequations}
which leads to the solution of $G(\phi)$ at the horizon as
\be
G(\phiH)=-\frac{4}{3}\, \frac{V(\phiH)}{V'(\phiH)}\,.
\ee
So, the near horizon solution of $G$ can be read as
\be
G(\phi) = -{4 \over 3}{V(\phi_{H}) \over V'(\phi_{H})} -
    {2 \over 3} \left( 1-\frac{V(\phi_{H})V''(\phi_{H})}{V'(\phi_{H})^{2}} \right)
    (\phi-\phi_{H}) + O\left[ (\phi-\phi_H)^2 \right] \, .
\ee
On the other hand, near boundary ($\phi\rightarrow0$) solution of $G$ turns out to be
\be
\label{eq:G0}
G(\phi)= \frac{1}{\Delta -4} \frac{1}{\phi} + \cdots \,,
\ee
where $\Delta$ is the dual operator scaling dimension which is 3 for the choice of potential \eqq{eq:pot} as mentioned above. 
Substituting the \eqq{eq:G0} into the \eqq{eq:A} and requiring that $A$ must be finite at the boundary, the function $A$ is obtained
\be
A(\phi) \simeq \frac{1}{\Delta -4} \, \log(\phi)\, .
\ee
Comparing this equation with \eqq{eq:A} at the limit of $\phi_0\rightarrow \phiH$ lead to 
\be
\label{eq:AH}
A_H \equiv A(\phiH)=\log \left(\frac{\phiH}{\Delta - 4 }\right) + \int_0^\phiH d\phi \left( G(\phi) - \frac{1}{(\Delta - 4)\phi }\right) \, ,
\ee 
and finally $A$ can be expressed as
\be
\label{eq:Af}
A(\phi)=A_H+ \int_\phiH^\phi d\phit \, G(\phit)  \, .
\ee 
There is a residual gauge freedom in the ansatz of metric $dr=\pm \,e^B\,d\phi $. As $\phi$ increases from zero to $\phiH$,  the radial coordinate $r$ decreases from $\infty$ to a finite value. Therefore the minus sign is meaningful. 
The leading asymptotic behavior of metric functions at large $r$  is
\be
 \label{FarAsymptotics}
  A \approx {r \over L} \qquad
  h \approx 1 \qquad
  \phi \approx (\Lambda L)^{4-\Delta} e^{(\Delta-4)A}\,.
 \ee
Therefore, the asymptotic behavior of generating function at large $r$ becomes
 \be
 \label{SmallPhiHandle}
  G = {dA \over d\phi} = {dr \over d\phi} {dA \over dr}
    \approx -e^B {1 \over L} \,.
 \ee
By considering the lower limit of the integral in the \eqq{eq:B} to be $\phiH$ and manipulating the equation, one can find the following relation
\be
\label{Bdiff}
  B(\phi) - B(\phi_H) = \log {G(\phi) \over G(\phiH)} +\frac{2}{3}
    \int_{\phi_H}^\phi {d\phit \over G(\phit)} \,.
\ee
Substituting $B(\phi)$ from \eqq{SmallPhiHandle} into the above equation, one can read $B(\phiH)$
\be
\label{BH}
B_H \equiv  B(\phiH) = \log (-G(\phiH)) +\frac{2}{3}
    \int_0^{\phiH} {d\phi \over G(\phi)} \,,
\ee
and finally we get
\be
\label{Bf}
B(\phi) =B_H  +  \int_{\phiH}^{\phi} d\phit \frac{G'(\phit)+2/3}{G(\phit)}\,.
\ee
In the last step, one can obtain the function $h(\phi)$ from \eqq{eq:phiphi} and \eqq{eq:VV} as follows
\be
\label{eq:h}
h(\phi)=-\frac{e^{2 B(\phi )} L^2 \left(4V(\phi )+3 G(\phi )V'(\phi )\right)}{3 G'(\phi )} \,.
\ee
The thermodynamical quantities of the background are expressed as
\be
L\,T= \frac{A(\phiH )-B(\phiH)}{4 \pi}\,,   \quad \quad  s= \frac{2 \pi}{\kappa_5^2} e^{3 A(\phiH)} \,.
\ee
Substituting the metric functions at the horizon leads to the following form for the temperature and entropy of the plasma
\bea
T&=&-\Lambda \frac{L^2 V(\phiH)}{3 \pi \phiH} \exp \left\{ \int_0^\phiH d\phi \left( G(\phi) + \frac{1}{\phi} +\frac{2}{3 G(\phi)}  \right)\right\} \, ,
\\[2mm]
s&=& \frac{2 \pi}{\kappa^2_5}  \frac{\Lambda^3}{ \phiH^{3}} \exp\left\{ 3 \int_0^\phiH d\phi  \left(G(\phi) + \frac{1}{\phi}\right)\right\} \,. 
\eea
\begin{figure}
\begin{subfigure}{.5\textwidth}
  \centering
  \includegraphics[width=0.95\linewidth]{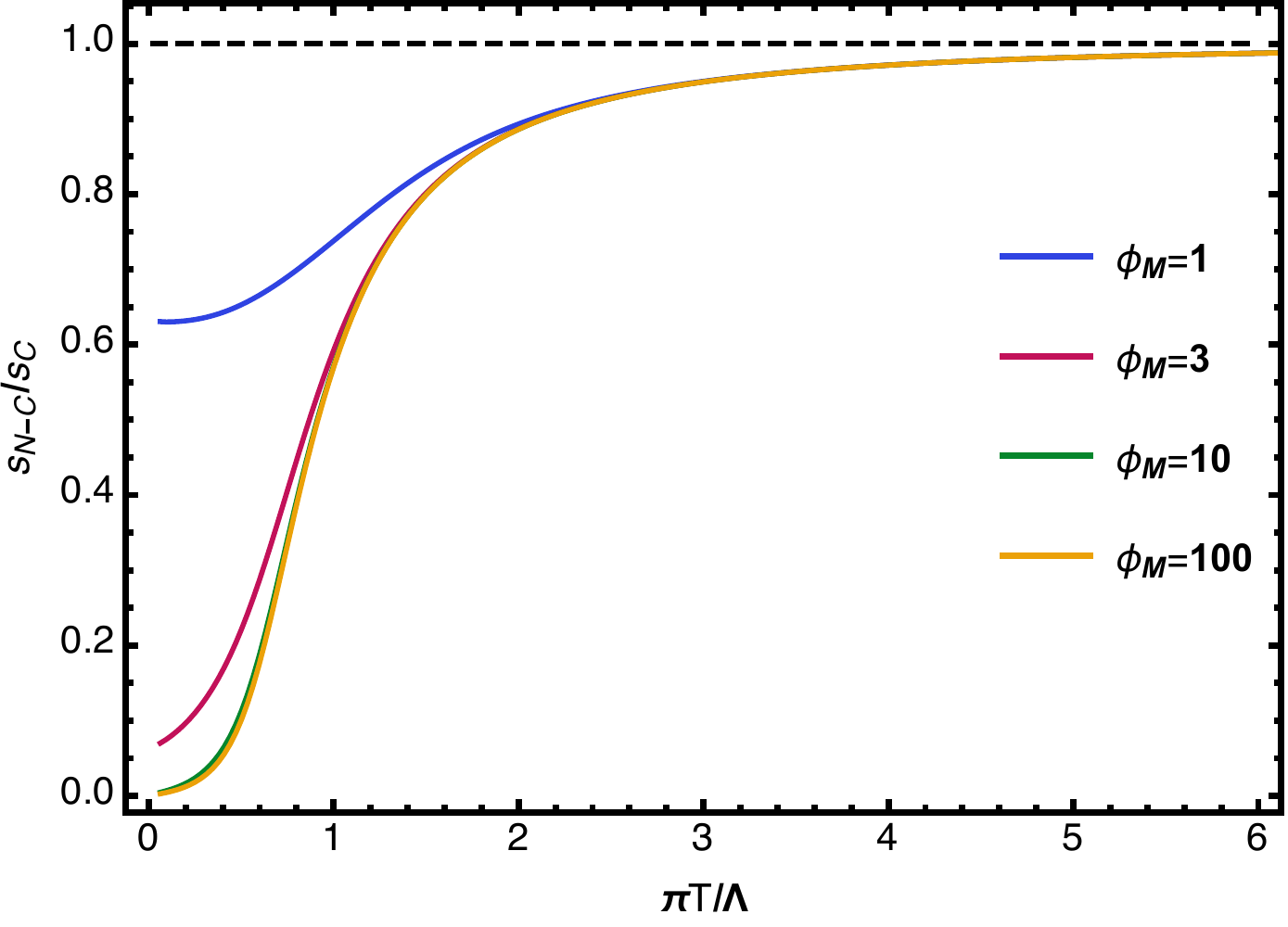}
   \caption{}
  \label{sratio}
\end{subfigure}%
\begin{subfigure}{.5\textwidth}
  \centering
  \includegraphics[width=0.97\linewidth]{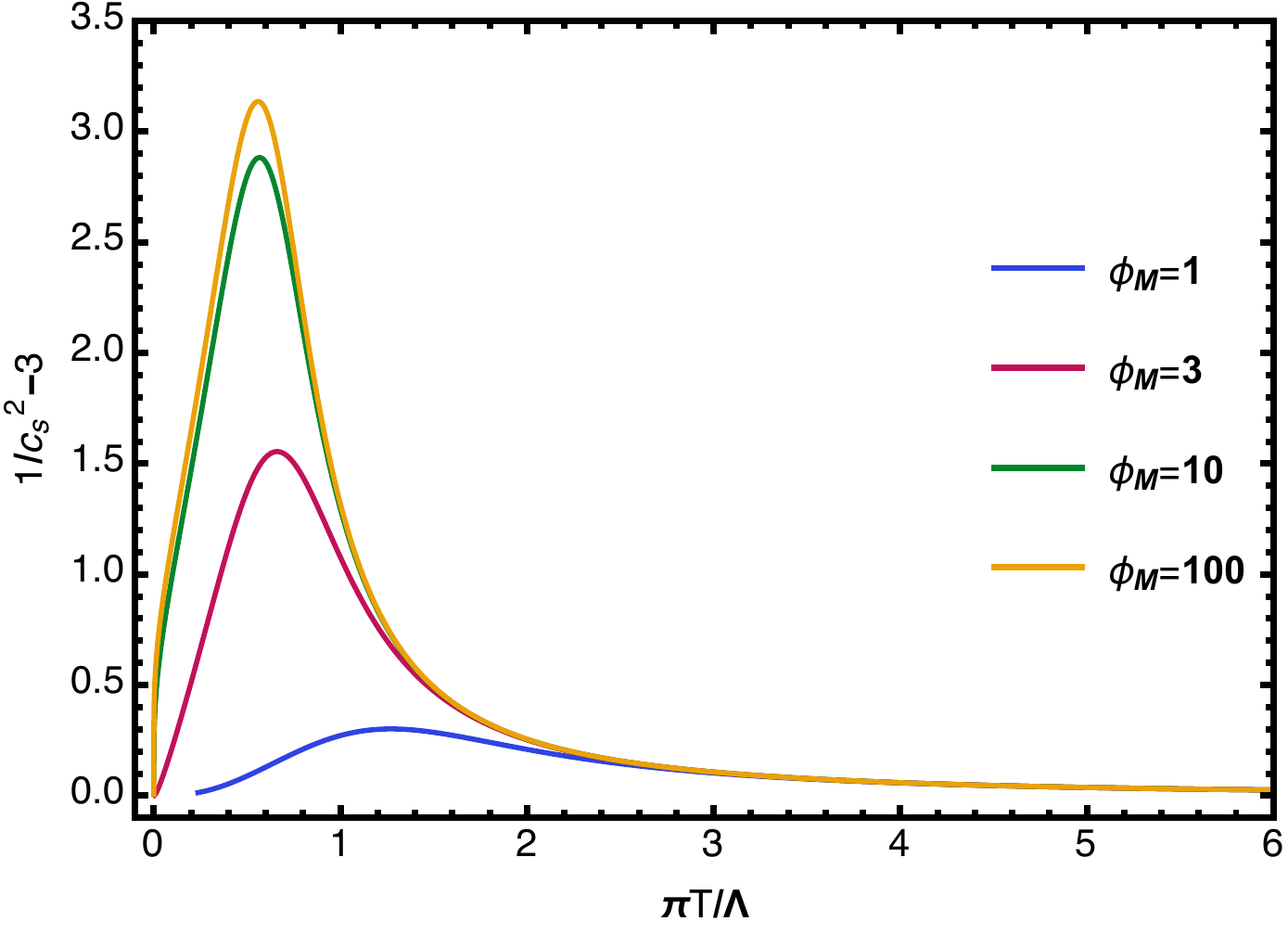}
   \caption{}
  \label{soundspeed}
\end{subfigure}%

\begin{subfigure}{.49\textwidth}
  \centering
  \includegraphics[width=0.97\linewidth]{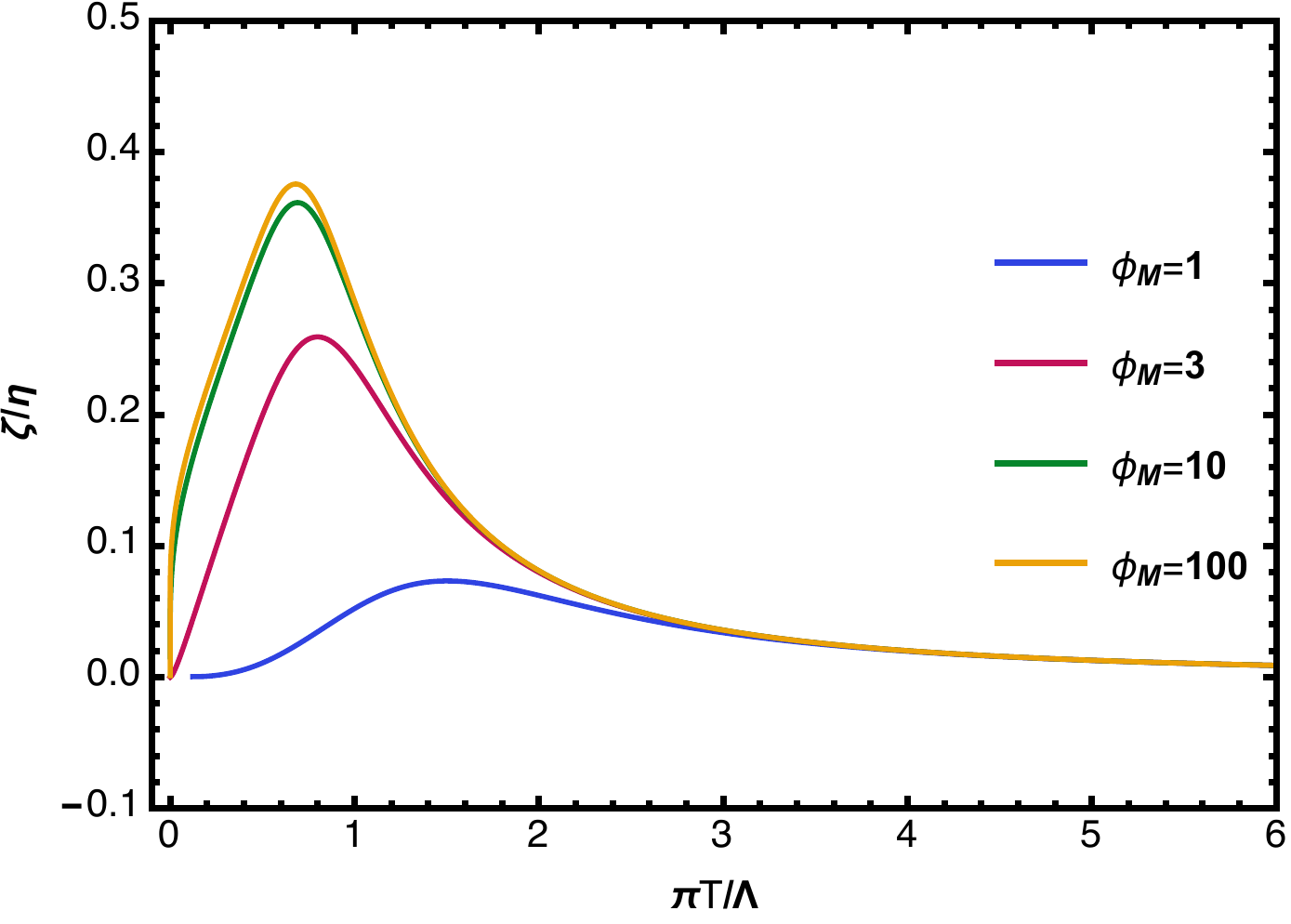}
  \caption{}
  \label{viscosity}
\end{subfigure}
\begin{subfigure}{.51\textwidth}
  \centering
  \includegraphics[width=0.99\linewidth]{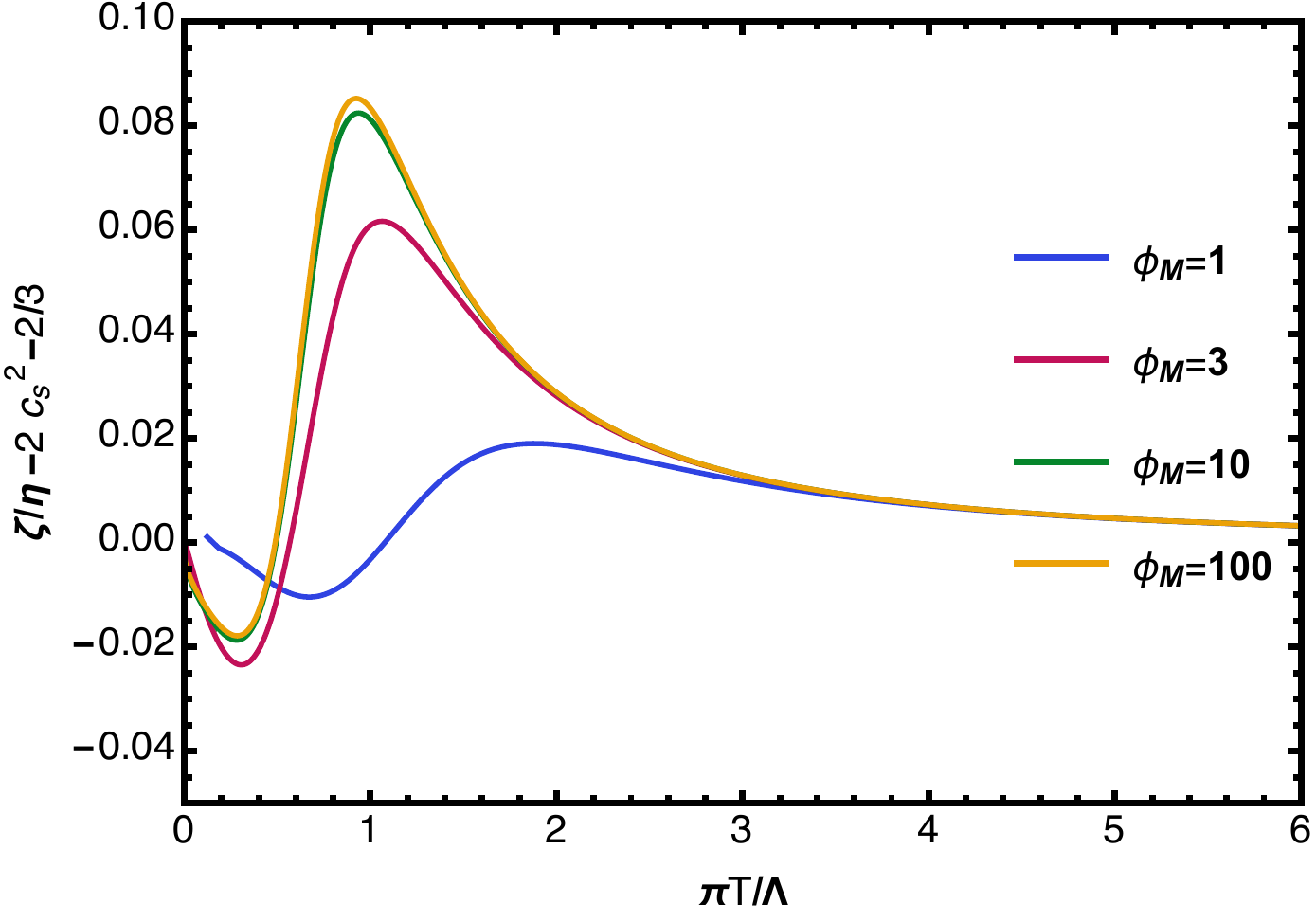}
  \caption{}
  \label{Buchelbound}
\end{subfigure}
\caption{(Color online) (\subref{sratio}) The ratio of non-conformal to the conformal entropy, (\subref{soundspeed}) inverse speed of sound square,  (\subref{viscosity}) ratio of bulk to shear viscosity, (\subref{Buchelbound}) violation of Buchel's bound as a function of temperature for different values of $\phiM$.}
\label{backgroundsplot}
\end{figure}

The non-conformal behavior of the thermodynamics of the dual theory can be identified by studying the entropy, the speed of sound, the ratio of bulk viscosity to shear viscosity and the Violation of Buchel’s bound. In \fig{sratio}, we plot the ratio of non-conformal to the conformal entropy in terms of temperature. At high and low temperature, the entropy of the theory is the same as the conformal theory and scale as $T^3$. 

The square of the speed of sound can be obtained from the inverse of the logarithmic derivative of the entropy as
\be
\frac{1}{c_s^2}=\frac{d\log s}{d \log T} \, .
\ee
The deviation of the speed of sound from its conformal value ($c_s=1/\sqrt{3}$) is depicted in the  \fig{soundspeed} as a function of temperature. At low and high temperatures, this quantity reaches its conformal value while the deviation gets larger by increasing $\phiM$ at intermediate temperatures.
 The ratio of shear viscosity $\eta$ to entropy has a universal value of $\eta/s=1/4\pi$ for all theories with a two-derivative gravity dual \cite{Kovtun:2004de}. The bulk viscosity, $\zeta$ can be determined from the dependency of the entropy on the value of the scalar field at the horizon \cite{Mas:2007ng}
\be
\frac{\zeta}{\eta}=4 \left(\frac{d \log s}{d\phiH } \right)^{-2} \, ,
\ee
which is zero in any conformal field theory.  This ratio has been shown in the  \fig{viscosity} as a function of temperature for different values of $\phiM$. The behavior of this quantity as a function of temperature is similar to that of the speed of sound. At low temperatures, the violation of this ratio from Buchel’s bound is shown in the \fig{Buchelbound}.

\section{Light quark maximum stopping distance}
\label{section:LightJet}

We are interested to study the propagation of energetic excitations which resemble quark jets through the strongly coupled non-conformal plasma. These excitations are considered as open string configurations with high energy, moving through the bulk of AdS space. The maximum stopping distance of these strings, the quantity which is not sensitive to the precise
initial conditions, can be used as a phenomenological guideline to estimate the stopping power of the strongly- coupled plasma. 

In this section, we first briefly review the string configuration dual to a light quark jet in the non-conformal background and then numerically calculate the maximum stopping distance which a quark with initial energy $E$ can travel.

\subsection{String configuration}
\label{section:FallingStrings}

\begin{figure}
\center
\includegraphics[scale=0.17]{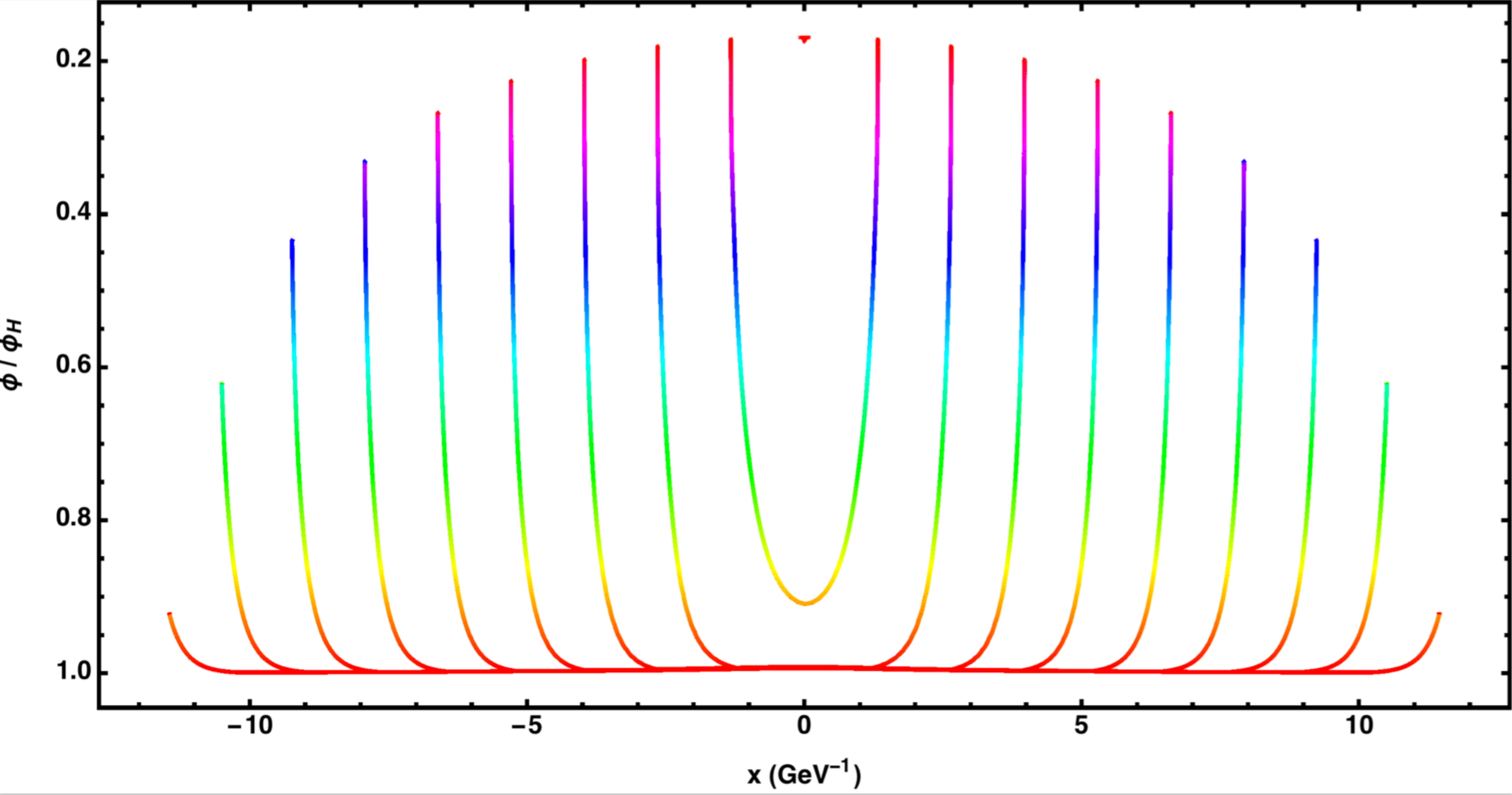}
\caption {The typical falling string profiles moving in the $x-\phi$ plane in the non-conformal background \eqq{eq:blackads} with $\phi_M=3$ and $T=0.265 \, GeV$. The string is created at a point at $\phi_c = 0.2$ and evolve to the extended object by time evolution.  The quark, corresponding to the half of the string, has $100$ GeV energy and  the small positive virtuality as defined in \eqq{virtuality}.
 \label{profile}}
\end{figure}

We use the black brane solution of  \eqq{eq:blackads} in Poincare coordinates to study the non-conformality effect on the dynamics of light quark jet moving in a strongly coupled plasma. According to the gauge/gravity duality,  quarks moving through the medium are dual to open strings moving in the 10-d gravitational background. D-branes are added to the background in the probe limit to describe the fundamental matter of the field theory such that do not back-react on the background metric \eqq{eq:blackads}. These branes fill the $4D$ Minkowski space and extend along the radial coordinate from the boundary at $\phi=0$ down to the location of the horizon at $\phi=\phi_H$. \\
According to the AdS/CFT dictionary, quark--antiquark pairs in the field theory side are dual to open strings whose endpoints are attached to the D-brane. In this paper, we consider a back-to-back jet pair created in the medium which can be regarded as a string created at a point, extends in space-time and falls toward the horizon. The profile evolution of a typical string for various times has been shown in \fig{profile}. In this setup, the two endpoints of the string move away from each other and the total spatial momentum of the string vanishes. The string embedding function is $X^{\mu}(\tau,\sigma) \mapsto \left( t(\tau,\sigma),x(\tau,\sigma),0,0,\phi(\tau,\sigma) \right) $ and for an open string created in a point $\phi=\phi_c$ at a time $t_c$, the profile is given by 
\be
\label{eq:emb}
    t(0,\sigma)=t_c\, ,\,\,\, x(0,\sigma)=0\, ,\,\,\, \phi(0,\sigma)=\phi_c ,
\ee
where $t_c$ is the time for which the string becomes an extended object from a point and $\sigma \in [0,\pi]$.\\
The Nambu-Goto action  becomes singular for falling string at late times, therefore we use the Polyakov action in which additional degrees of freedom are being involved by a nontrivial worldsheet metric $\eta^{ab}$  and the equations of motion are well-behaved everywhere on the worldsheet \cite{Herzog:2006gh,Chesler:2008wd,Chesler:2008uy}. The Polyakov action for the string has the following form
\be
\label{Polyakov action}
    S_P=-\frac{T_0}{2}\int d^2\sigma \> \sqrt{-\eta} \, \eta^{ab}\,
    \partial_a X^\mu\partial_b X^\nu \, G_{\mu\nu}\, .
\ee
Canonical momentum densities associated with the string can be obtained by varying the action with respect to the derivatives of the embedding functions, 
\be
\label{Canonical momentum}
    \Pi^a_{\mu}(\tau,\sigma)
    \equiv
    \frac{1}{\sqrt {-\eta\,}} \,\frac{\delta S_{\rm P}}{\delta ( \partial_a X^\mu(\tau,\sigma))} = -T_0\, \eta^{ab}\, \partial_b X^\nu\, G_{\mu  \nu},
\ee
and the equations of motion are being obtained by variation of the Polyakov action with respect to the embedding functions $X^\mu$ as
\begin{align}
	\label{stringEoM}
	\partial_{a} \big[ \sqrt{-\eta}\,\eta^{ab} \, G_{\mu\nu}\, \partial_{b}X^{\nu}  \big] & = \frac{1}{2} \sqrt{-\eta}\,\eta^{ab}
    \frac{\partial G_{\nu \rho}}{\partial X^{\mu}} \,
    \partial_{a}X^{\nu}\partial_{b}X^{\rho} \nonumber \\[5pt]
    \Longleftrightarrow \qquad \nabla_a\,  \Pi^a_{\mu} & = -\frac{T_0}{2} \,\eta^{ab}
    \frac{\partial G_{\nu \rho}}{\partial X^{\mu}} \,
    \partial_{a}X^{\nu}\partial_{b}X^{\rho}.
\end{align}
We choose the following worldsheet metric
\be
\label{Worldsheet Metric}
    \|\eta_{ab}\|= \left(\begin{array}{cc} -\Sigma(\phi) & 0 \\ 0 & \Sigma(\phi)^{-1}
    \end{array}\right),
\ee
where $\Sigma$ is called the stretching function and generally can be an arbitrary function of worldsheet embedding functions. We found that choosing the stretching function of the form
\be
\label{eq:stretch}
\Sigma(\phi)=
\left(\frac{e^{A(\phi)}}{e^{A(\phi_c)}}\right)^a
\left(\frac{h({\phi)}}{h({\phi_c)}}\right)^b
\ee
would cancel the singularities and equations of motion remain well behaved everywhere on the worldsheet. We choose values of $a$ and $b$ in the range of 1 to 5 depends on the string initial conditions.  The constraint equation is obtained by variating the Polyakov action, \eqq{Polyakov action} with respect to $\eta^{ab}$ which for a string with point-like initial condition reduces to $\dot{X}^2(0,\sigma)=0$ at initial time. Combining the open string boundary condition and constraint equation, the initial profile must satisfy the following condition
\be
\label{finalBC}
       \dot{x}'(0,\sigma^*)=\dot{ \phi}'(0,\sigma^*)=0.
\ee

The below initial condition (IC) obeys all necessary conditions and assures the physical requirements for the non-conformal background
\begin{eqnarray}
   &&  \dot{x}(0,\sigma) =A \,  \phi_{\rm c} \cos \sigma\,, \nonumber \\
    && \dot{ \phi}(0,\sigma) =  \phi_{\rm c}\ (1-\cos 2\sigma)\,, \\
   && \dot{t}(0,\sigma) = \sqrt{\frac{g_{xx}(\phi_c)\,\dot{x}^2+g_{\phi\phi}(\phi_c)\,\dot{\phi}^2}{-g_{tt}(\phi_c)}}\, ,\label{eq:IC} \nonumber
 \end{eqnarray}
where $\phi_c$ and $A$ are free parameters related to the energy and momentum of the dual quark in the field theory side. The IC are chosen such that the string is long-lived and yields stable numerical solutions.  Also, most of the string's energy and momentum are concentrated near its endpoints. These IC yield a symmetric string profile about $x=0$ at all times, because $\dot{x}(0,\sigma)$ is antisymmetric about $\sigma=\pi/2$ while $\dot{\phi}(0,\sigma)$ is symmetric. We performed the explicit check of any numerical solution, and our solutions respected the equation of constraint for all $\tau$.

Now, we proceed to finding the light quark  energy and momentum. As the non-conformal background geometry $G^{\mu\nu}$ depends only on $\phi$, hence for $\mu$ corresponding to $(t,\,\vec{x})$ one can write
\be
\nabla_a\Pi_\mu^a = 0 \, . 
\label{covariant EQM}
\ee
Therefore, the momentum densities $\Pi^a_\mu$ are conserved Noether currents on the related worldsheet. The total energy of the falling string in the non-conformal geometry of \eqq{eq:blackads} at any time is constant and equal to the initial energy of the string
\begin{eqnarray}
E_{\rm string}  &=& - \int_0^\pi d \sigma \> \sqrt{-\eta} \, \Pi^\tau_{t}(0,\sigma) \nonumber \\
&=& \frac{\lambda}{2 \pi} \int_0^\pi d \sigma \> \, \sqrt{{g_{tt}(\phi_c)\left(g_{xx}(\phi_c)\,{\dot x(0,\sigma)}^2+g_{\phi\phi}(\phi_c)\,{\dot \phi(0,\sigma)}^2 \right)} }\, ,   \nonumber \\
\end{eqnarray}
and the total energy of the light quark is obtained from
\be
E_q=\frac{1}{2} E_{\rm string} \,.
\ee
The total momentum of the string is conserved in $x$-direction and can be find from the string IC as
\be
P_{q} =  \frac{\lambda}{2 \pi} \int_0^{\pi/2} d \sigma \> \, g_{xx}(\phi_c)\,{\dot x(0,\sigma)} \, .
\ee
Here, we have considered that the total momentum of the string in $x$-direction is completely symmetric about $\sigma=\pi/2$ and equals to zero . Therefore, the quark momentum equals to the anti-quark momentum with a minus sign. Also, the string energy and momentum are determined by two parameters $A$ and $\phi_c$ in the string initial condition. These parameters are translated into the virtuality of the jet created in the non-conformal QGP which is obtained from
\be
Q^2 \equiv E_q^2 - P_q^2 
\label{virtuality}
\,.
\ee
The solution of \eqq{stringEoM} for a typical string with $E=100$ GeV in the non-conformal background \eqq{eq:blackads} with $\phi_M=3$ is plotted in \fig{profile}. The temperature of the plasma is about $T=0.265\, GeV$. The dynamics of the two half of the string are exactly the same, one moving in the opposite direction of the another one. As we expected, the two endpoints of the string move away from each other as the string extends along the $(x, \phi)$ direction and falls toward the horizon. 

\subsection{Maximum stopping distance}
\label{section:Stoppingdistance}
 
\begin{figure}
\center
\includegraphics[scale=0.5]{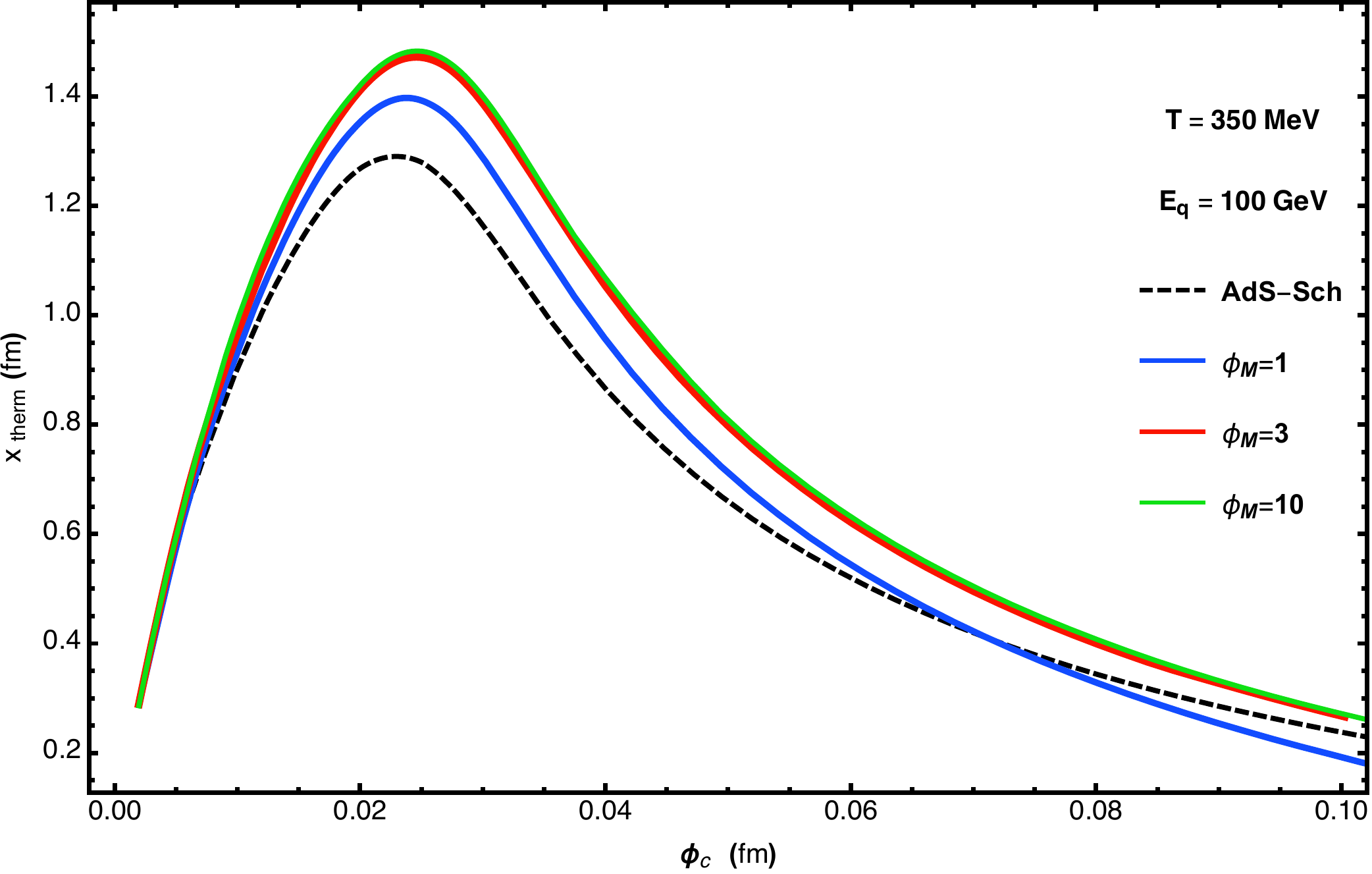}
\caption {(Color online) 
The maximum stopping distance of the jet moving in the plasmas versus to its initial distance from the boundary located at $\phi=0$. The non-conformal plasma has the same temperature as the conformal plasma, $T=$ 350 MeV. 
In all cases, the quark has 100 GeV energy but created at a different radial distance from the boundary. The dashed black line is the stopping distance in the AdS-Sch background, while the blue, red and green lines are stopping distance in the non-conformal plasmas with $\phi_M=$ 1, 3 and 10, respectively. The stopping distance increases by increasing $\phi_M$ in the medium ranges of $\phi_c$ where it is maximum. 
\label{fig:Xtherm}}
\end{figure}

Numerical studies, \fig{fig:Xtherm} show that the distance which a quark with a specific energy traversed in the plasma before thermalizing highly depends on the chosen initial condition of the dual string, i.e. the initial distance from the boundary. Since there is no known map between the string IC and the actual field theory quantities, we aim to study a quantity which is insensitive to these IC in the string side. The stopping distance which is called the thermalization distance, $x_{therm}$ is defined as the length along the x-direction from the point of production of the original point-like string to the point at which the end of the string falls through the black hole horizon. On the field theory side of the duality, $x_{therm}$ corresponds to the length of the plasma traversed before the jet becomes completely thermalized (i.e. indistinguishable from the plasma). We are looking for the maximum distance which a quark with energy $E$ can travel before thermalizing independent of the exact IC of the string. Although this quantity is not enough to calculate observables such as $R_{AA}$ or $v_2$ in general, it can be used as a phenomenological guideline to estimate the stopping power of the strongly- coupled plasma. 

\begin{figure}
 \center
  \includegraphics[width=0.88\linewidth]{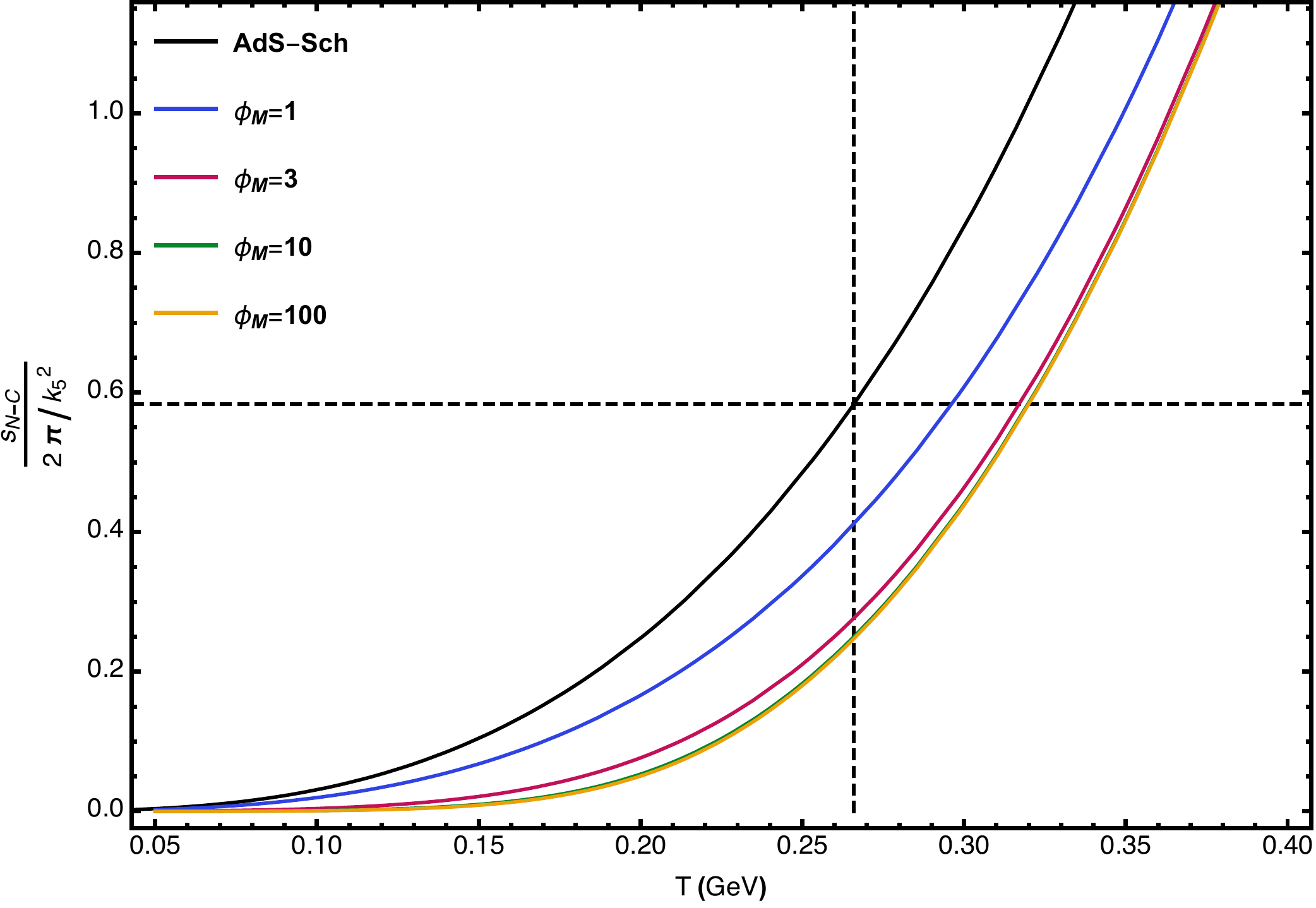}
  \caption{ (Color online) The rescaled entropy of the non-conformal theory with different $\phi_M$ compared to the conformal entropy in terms of temperature. The intersection of two dashed black lines shows an AdS-Sch background with $T=265$ MeV. We study the non-conformality effect on our results by moving away from this point along the either vertical (same temperature) or horizontal (same entropy) dashed black lines.}
\label{fig:entropy}
\end{figure}

In order to compare the results of stopping distance of jet in non-conformal plasma with the results of conformal AdS-Sch, we can either fix the temperature or the entropy of the non-conformal plasma with the AdS-Sch ones. In \fig{fig:entropy} we plot the rescaled entropy of the non-conformal theory with different $\phi_M$ compared to the conformal entropy in terms of temperature.  The intersection of black dashed lines which shows the entropy of the AdS-Sch metric with $T=265$ MeV is our reference point. This figure indicates that if we fix the temperature and increase $\phi_M$ (moving along the vertical dashed black line), the entropy of the non-conformal theory decreases. On the other hand, increasing $\phi_M$ on the line of fixed entropy (moving along the horizontal dashed black line) give rises to the non-conformal theories with higher temperature. 

\begin{figure}
\center
\includegraphics[scale=0.45]{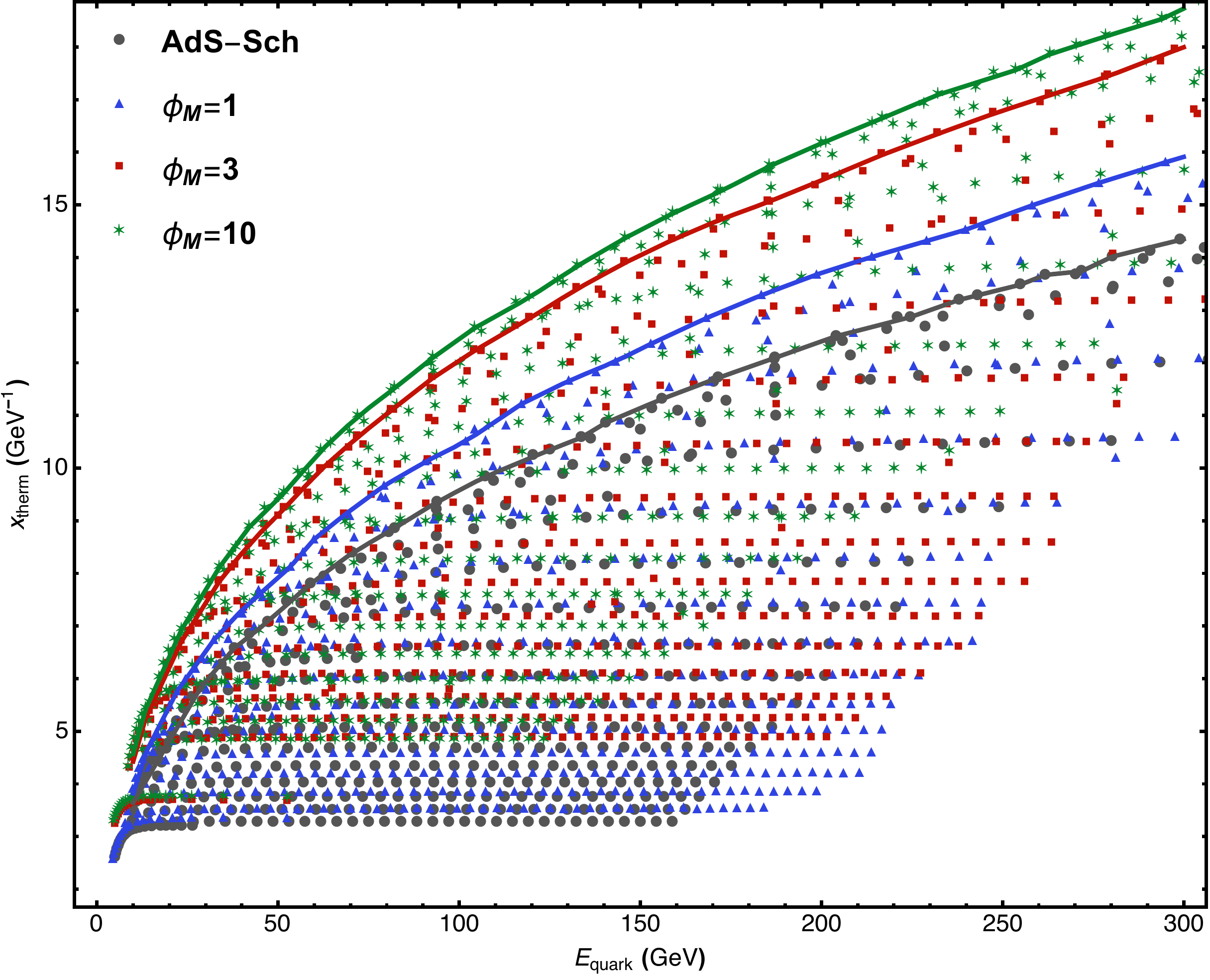}
\caption {(Color online) The stopping distance of the jet in terms of its energy moving in the non-conformal backgrounds with different $\phi_M$. All of the non-conformal backgrounds have the same temperature as the AdS-Sch temperature sets to be 265 MeV. For each set of data, all data points fall below the solid line which determines the maximum stopping distance of the quark in terms of its energy. 
 \label{fig:StoppingDT}}
\end{figure}

\begin{figure}
\center
\includegraphics[scale=0.45]{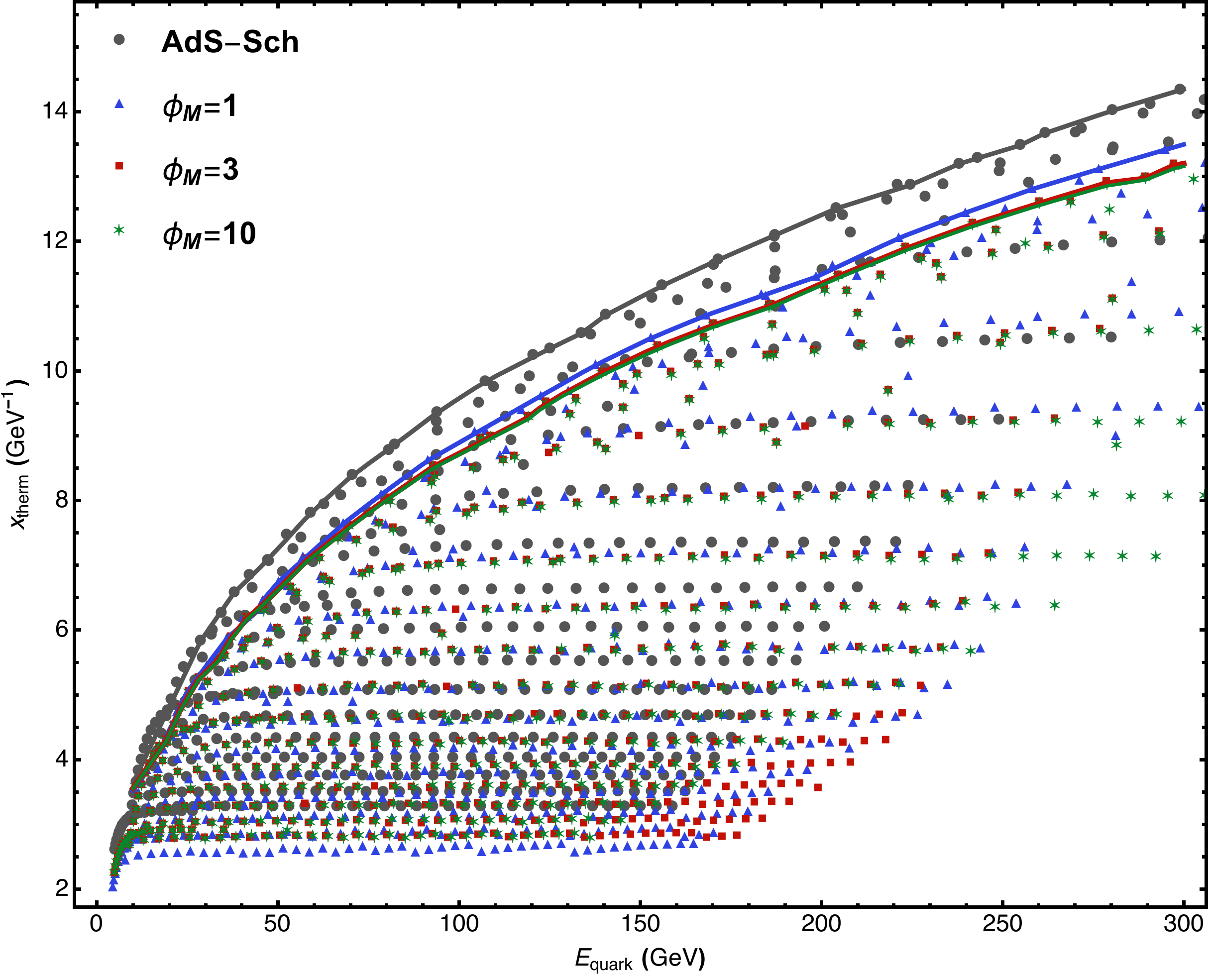}
\caption {(Color online) 
 The stopping distance of the jet in terms of its energy moving in the non-conformal backgrounds with different $\phi_M$. All of the non-conformal backgrounds have the same entropy as the AdS-Sch entropy. For each set of data, all data points fall below the solid line which determines the maximum stopping distance of the quark in terms of its energy. 
 \label{fig:StoppingDS}}
\end{figure}

In \fig{fig:StoppingDT}, and  \fig{fig:StoppingDS} we numerically compute the stopping distance for many different sets of string initial conditions in backgrounds with different non-conformal parameter. In these figures, each point shows the stopping distance of a string with a specific initial condition in terms of its initial energy. Black dots show the strings in the AdS-Sch metric, while the blue triangles, red squares, and green stars correspond to strings in the non-conformal backgrounds with $\phi_M=1$, $\phi_M=3$, and $\phi_M=10$, respectively.  In the \fig{fig:StoppingDT}, all plasmas have the same temperature while in the \fig{fig:StoppingDS} the entropy of the backgrounds is fixed. The interesting point is that the stopping distance increases with increasing the deviation from conformality when the temperature of the background is fixed while it decreases if the entropy of the background is fixed. It is expected, since as explained in \fig{fig:entropy}, by fixing the entropy and increasing the $\phi_M$ the plasma becomes hotter which leads to a larger jet suppression.


As clearly shown in these figures, the dynamics of the string is a multi-variable quantity depends on both string IC and background thermodynamics. But, all data points for each set of data fall below the solid lines. These lines display the maximum stopping distance of the light quark in the corresponding medium, i. e. the maximum distance that a quark with energy E can travel in the medium before thermalization. It is shown that this distance in the conformal background scales like $E^{1/3}$ for enough large energies \cite{Chesler:2008uy}. In fact, numerical results show that the maximum stopping distance for a given energy depends on the energy of the quark and temperature of the plasma as follows 
\be
x_{max}=\frac{\mathcal{C}}{T}(\frac{E}{T\, \sqrt{\lambda}})^{n_{eff}}
\label{eq:xtherm}
\,,
\ee
which $\mathcal{C}$ and $n_{eff}$ is estimated to $0.526$ and $1/3$ in the AdS-Sch case, respectively \cite{Chesler:2008uy}. We explore this relation with our numerical results and estimate the values of $\mathcal{C}$ and $n_{eff}$ in the case of non-conformal plasma. We fit our data in the AdS-Sch with \eqq{eq:xtherm} by fixing $\mathcal{C}=0.526$. The estimated $n_{eff}$ is less than $1/3$ for the available energies at RHIC or LHC but it approaches this value at very high energies. 

\begin{figure}
\begin{subfigure}{.5\textwidth}
  \centering
  \includegraphics[width=.97\linewidth]{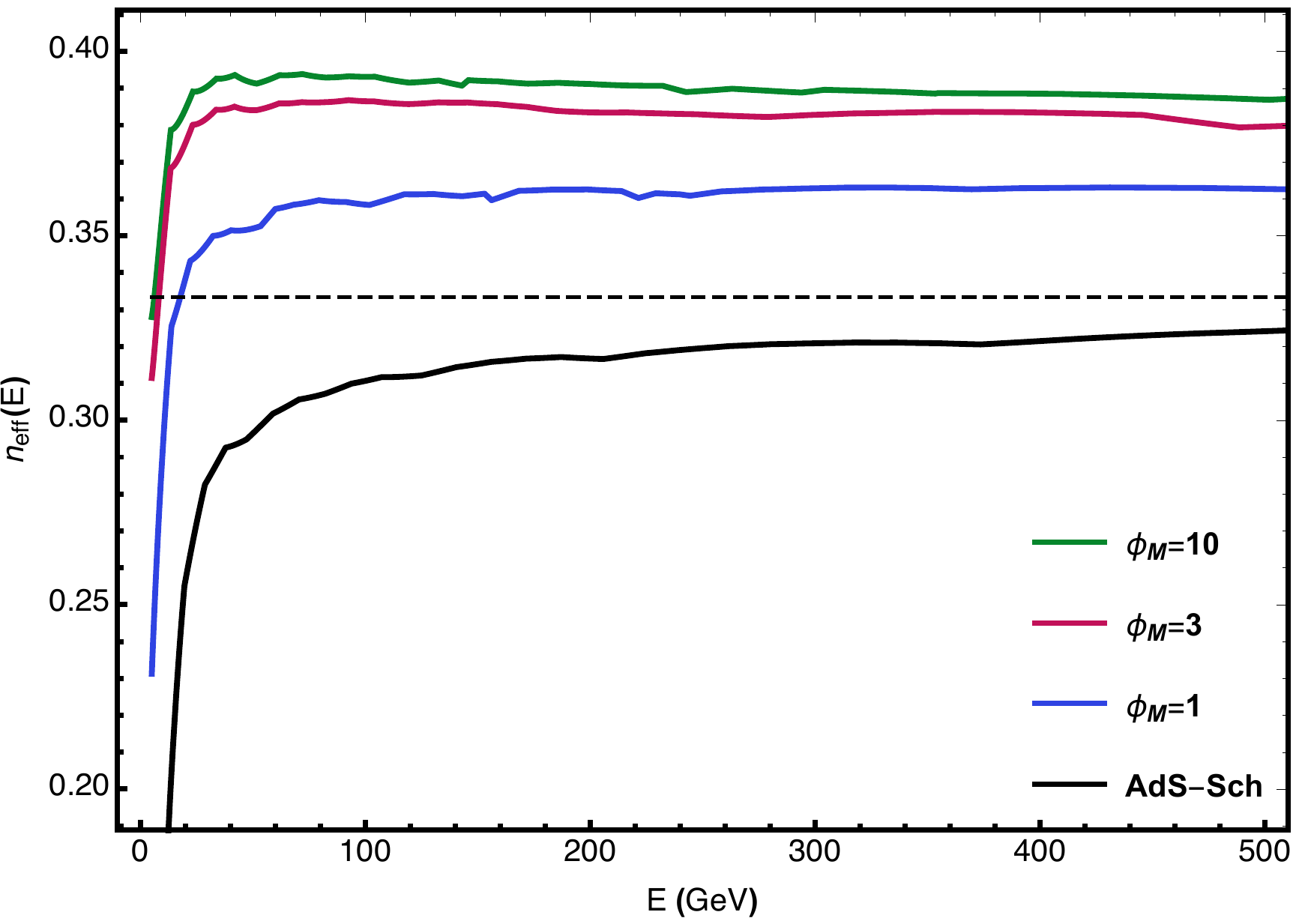}
  \caption{}
  \label{fig:neffT}
\end{subfigure}
\begin{subfigure}{.5\textwidth}
  \centering
  \includegraphics[width=.97\linewidth]{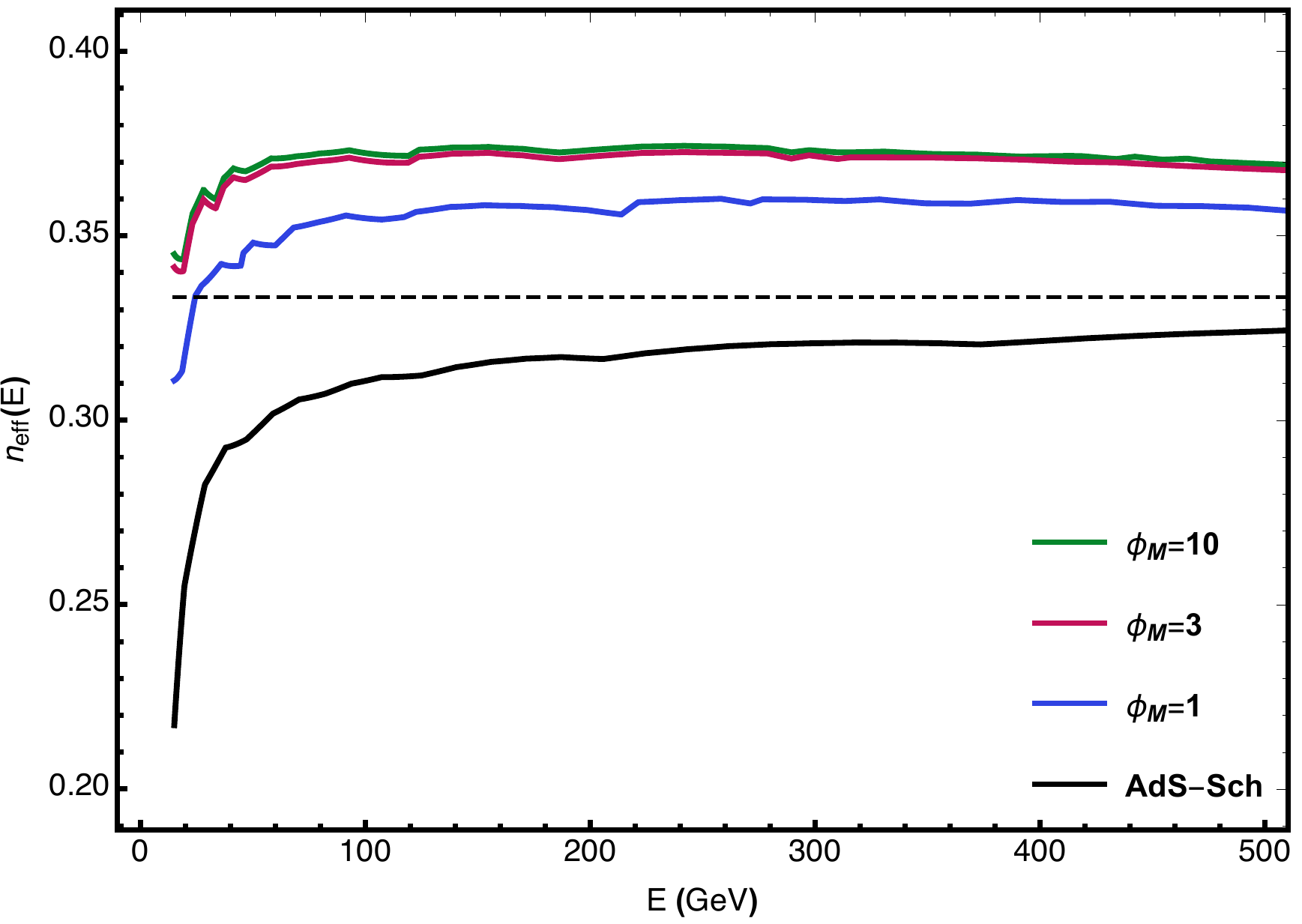}
  \caption{}
  \label{fig:neffS}
\end{subfigure}
\caption{(Color online) (\subref{fig:neffT})  The effective power in \eqq{eq:xtherm}  in non-conformal plasmas with the same temperature as the AdS-Sch temperature. (\subref{fig:neffS}) The effective power in \eqq{eq:xtherm} in non-conformal plasmas with the same entropy as the AdS-Sch entropy. 
}
\label{fig:neff}
\end{figure}

In \fig{fig:neff}, the $n_{eff}$ is calculated by fitting the numerical results to the \eqq{eq:xtherm} by considering the $\mathcal{C}=0.45$ for all non-conformal metrics and $\mathcal{C}=0.526$ for the AdS-Sch metric. In fact, these constants give rise to a more precise fit to the numerical results. Again, in \fig{fig:neffT} all backgrounds have the same T and in \fig{fig:neffS} they have the same entropy. In general, increasing the deviation from conformal invariance increases the stopping distance of jets of light quark.

\section{Jet Quenching}
\label{section:JQ}
\begin{figure}
\begin{subfigure}{.5\textwidth}
  \centering
  \includegraphics[width=.98\linewidth]{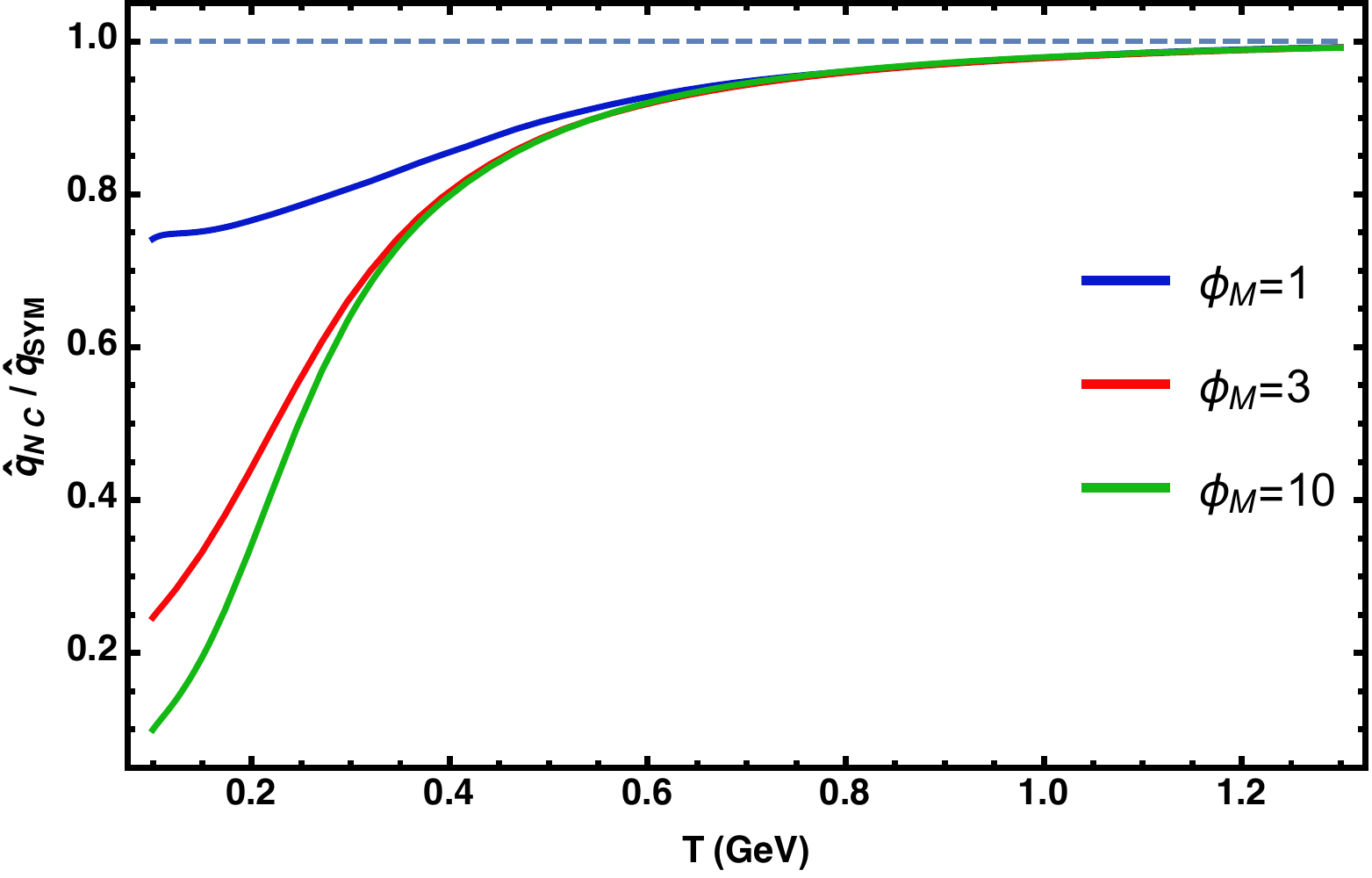}
  \caption{}
  \label{fig:qr}
\end{subfigure}
\begin{subfigure}{.5\textwidth}
  \centering
  \includegraphics[width=.98\linewidth]{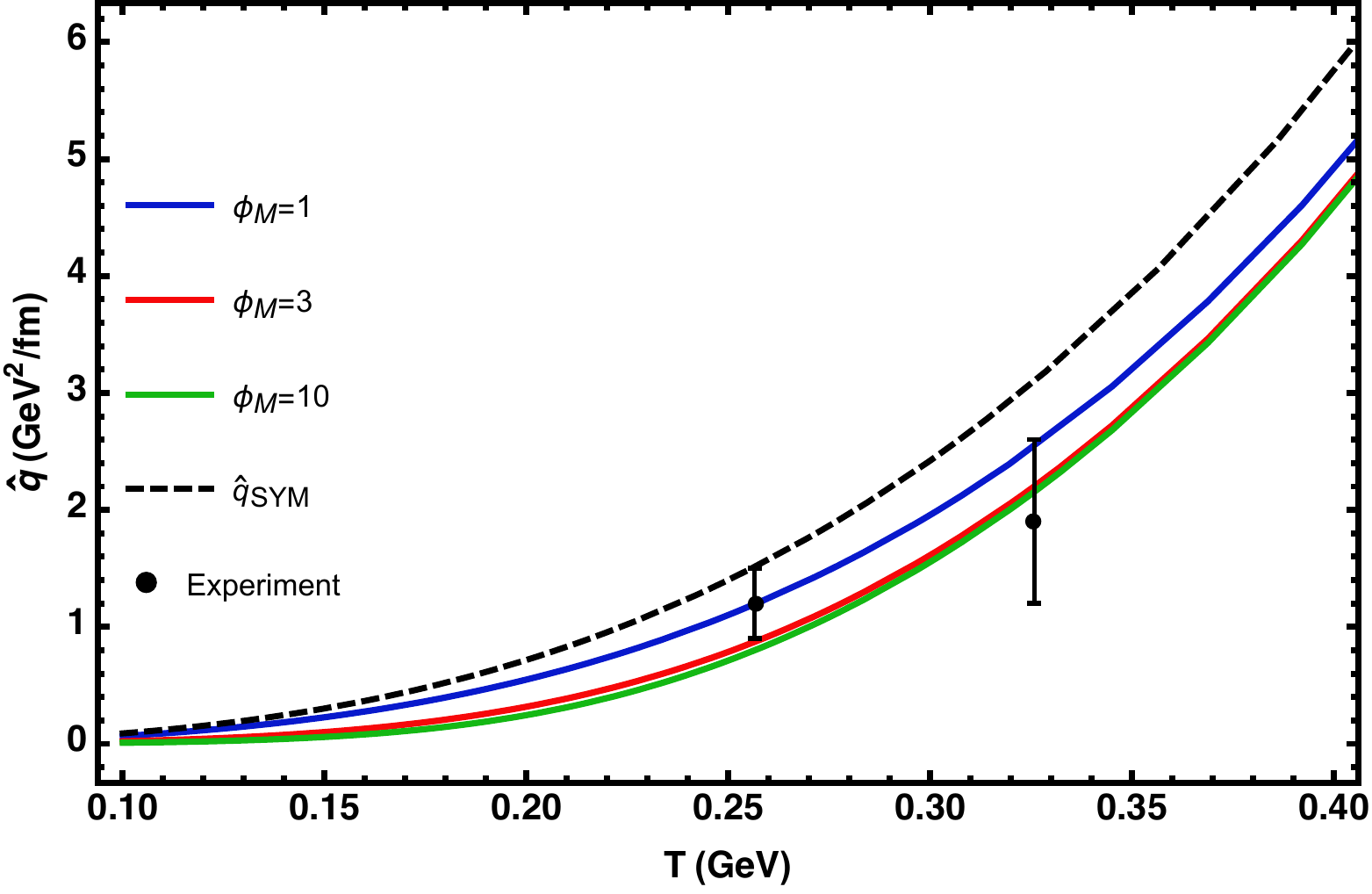}
  \caption{}
  \label{fig:q2}
\end{subfigure}
\caption{ (\subref{fig:qr})The ratio of $\hat{q}_{NC}/\hat{q}_{SYM}$ in terms of temperature for different $\phi_M$. (\subref{fig:q2}) Jet quenching parameter for $\mathcal N=4$ SYM theory and non-conformal theory with different $\phi_M$. The circles indicate the experimental values of $\hat{q}$ from RHIC and LHC.}
\label{fig:q}
\end{figure}

To calculate the jet quenching parameter for the non-conformal theory, we use the thermal expectation value of a close light-like Wilson loop \cite{Zakharov:1997uu}
\be
\langle{W^A({\cal C})}\rangle  \approx \exp \left[ - \frac{1}{4\sqrt{2}}  \hat{q}\, L^-\, L^2 \right]\, ,
\label{eq:wilson1}
\ee
where $L$ is the transverse distance (conjugate to the transverse momentum of the radiated gluons) and $L^-$ is the light-cone distance ( conjugate to partons with relativistic velocities). This equation is valid for  $L^-\gg L$.
In the gravity side, the thermal expectation value of the Wilson loop $\langle{W^F({\cal C})}\rangle$ can be obtained using the action of the extremal  surface as \cite{Maldacena:1998im,Rey:1998ik,Rey:1998bq,Brandhuber:1998bs,Sonnenschein:1999if}      
 \be
\langle{W^F({\cal C})}\rangle = \exp \left[ -  S_I({\cal C})\right] \, ,
\label{eq:wilson2}
\ee 
where $S_I$ is the normalized action after subtracting the self energy of the $q\bar{q}$ pair from the Nambu-Goto action of the string worldsheet. As in the large $N_c$ limit, ${\rm Tr}_{(Adj.)} = {\rm Tr^2}_{(Fund.)}$, one can write
\be
\hat{q}=\frac{8\sqrt{2}S_I}{L^-\, L^2}.
\label{eq:q-SI}
\ee 
In the light-cone coordinates, $x^{\pm}=(t\pm x_1)/\sqrt{2}$, the non-conformal background of  \eqq{eq:blackads} becomes
 \begin{eqnarray}
ds^2& = & -e^{2A}(1+h(\phi))dx^+ dx^- +e^{2 A}\left(d{x_2}^ 2 + d{x_3}^2 \right) 
 \nonumber \\
&&+\frac{e^{2 A}}{2}\left(1-h(\phi)\right) \left(d{x^+}^ 2 + d{x^-}^2 \right) +\frac{e^{2B}}{h} d\phi^2 \nonumber \\
 & \equiv & G_{\mu \nu} dx^\mu dx^\nu \, .
\label{eq:metric-l-c}
 \end{eqnarray}
The string parametrization is $x^{\mu}(\tau,\sigma)$ where $\sigma^\alpha=(\tau,\sigma)$ is the worldsheet coordinates. The Nambu-Goto action of the string is
 \be
S_{NG} ={1 \over 2 \pi \alpha'} \int d\sigma d \tau \, \sqrt{ \det g_{\alpha\beta}},
\label{eq:NG}
 \ee
where $g_{\alpha \beta}$ is the induced metric on the string worldsheet and $(\tau,\sigma)$ are set to be $(x^-,x_2)$. $L$ is considered to be the contour length along $x_2$-direction and $L^-$ to be its length along $\tau$-direction. The boundary conditions are $\phi(\pm\frac{L}{2})=0$ and $x_3(\sigma)$ and $x^+(\sigma)$ coordinates are constant. Then the action of \eqq{eq:NG} reads
\be
 S_{NG} = \frac{2L^-}{2 \pi \alpha'}  \int_0^{{L \over 2}} d \sigma \,
\sqrt{\frac{e^{4A}(1-h(\phi))}{2}}\sqrt{1+\frac{e^{2(B-A)}}{h(\phi)}\,\phi'^2}\ ,
\label{eq:NG2}
\ee
where prime denotes the derivative with respect to $\sigma$. Since the Lagrangian density is time independent, the Hamiltonian of the system is constant
\be
\mathcal{L}-\phi' \frac{\partial \mathcal{L}}{\partial \phi'}=\Pi_{\phi}
\label{eq:H}
\ee
and one can obtain the following equation for $\phi'$
\be
\phi' =\frac{\partial{\phi}}{\partial \sigma}=\frac{e^{A-B}}{\sqrt{2}{\Pi_{\phi}}} \sqrt{h(\phi)\left(e^{4A}(1-h(\phi))-2{\Pi_{\phi}}^2 \right)}.
\label{eq:phi}
\ee 
Integrating equation  \eqq{eq:phi} leads to
\be
\frac{L}{2} = \sqrt{2}a_0{\Pi_{\phi}} +\mathcal{O}({\Pi_{\phi}}^3), 
\label{eq:L}
\ee
where
\be
a_0=\int_{\phi_H}^{{0}} d \phi \frac{e^{B-3A}}{\sqrt{h(\phi)(1-h(\phi))}}\,.
\label{eq:a0}
\ee
Here, we have used the fact that for small length $L$, the constant $\Pi_{\phi}$ is small and its higher order terms are negligible. 
Substituting \eqq{eq:phi} into \eqq{eq:NG2} and expand for small $\Pi_{\phi}$ yields 
\be
S_{NG}=\frac{L^-}{\sqrt{2}\pi \alpha'}\int_{\phi_H}^{{0}} d \phi \sqrt{\frac{1-h(\phi)}{h(\phi)}}e^{A+B}\left(1+\frac{e^{-4A}\,{\Pi_{\phi}}^2}{1-h(\phi)}+...\right).
\label{eq:NG3}
\ee 
This action diverges and we should subtract the self energy of two disconnected strings whose worldsheets are located at $x_2=\pm\frac{L}{2}$ and extended from $\phi=0$ to $\phi=\phi_H$ 
\be
S_{0}=\frac{2L^-}{2\pi \alpha'}\int_{\phi_H}^{{0}} d \phi \sqrt{g_{--}g_{\phi\phi}}=\frac{L^-}{\sqrt2\pi \alpha'}\int_{\phi_H}^{{0}} d \phi \,e^{A+B}\sqrt{\frac{1-h(\phi)}{h(\phi)}}\,.
\label{eq:S0}
\ee 
The normalized action is then written as
\be
S_{I}=S_{NG}-S_0\equiv\frac{L^- L^2}{8\sqrt2\pi \alpha' a_0}\,.
\label{eq:SI}
\ee 
Inserting \eqq{eq:SI} into \eqq{eq:q-SI} leads to the following expression for the jet quenching parameter of the non-conformal theory
\be
\hat{q}_{NC}=\frac{\sqrt{\lambda}}{\pi R^2 a_0},
\label{eq:qfinal}
\ee 
where $a_0$ is the numerical integral of \eqq{eq:a0}. Here, we have used the relation between the string tension and the tÕHooft coupling $\frac{R^2}{\alpha'}=\sqrt{\lambda}$.\\
For $\mathcal N=4$ supersymmetric Yang-Mills theory in the large $N_c$ and large $\lambda$ limits, the jet quenching parameter is find to be \cite{Liu:2006ug}
\be
\hat{q}_{SYM}=\frac{\pi^{3/2}\Gamma(\frac{3}{4})}{\Gamma(\frac{5}{4})}\sqrt{\lambda} \,T^3,
\label{eq:qSYM}
\ee
To understand how non-conformality affects the jet quenching parameter, we calculate \eqq{eq:qfinal} for different  values of $\phi_M$ numerically. Here, we have taken the 't Hooft coupling $\lambda=5.5$ for numerical estimates. The ratio of jet quenching in non-conformal background to its conformal value in terms of temperature is shown in \fig{fig:qr} for $\phi_M=1$ (blue curve),  $\phi_M=3$ (red curve) and $\phi_M=10$ (green curve). The plot shows that for each value of $\phi_M$, the ratio of $\hat{q}_{NC}/\hat{q}_{SYM}$ starts from a finite value and approaches to 1 at higher temperature where conformality is dominant. This behavior reflects the fact that deviations from conformality are magnificent at lower temperatures and suppress at higher temperatures. Also, increasing the values of $\phi_M$, decreases the values of jet quenching (and hence the ratio). In \fig{fig:q2}, the temperature dependency of the jet quenching parameter is shown for $\phi_M=1$ (blue curve),  $\phi_M=3$ (red curve), $\phi_M=10$ (green curve) as well as $\mathcal N=4$ SYM theory (dashed curve). Two black circles with error bars are the corresponding absolute values for $\hat{q}$ for a 10 GeV quark jet in the most central Au-Au collisions at RHIC with highest temperature $T=0.37 GeV$ and Pb-Pb collisions at LHC with highest temperature $T=0.47 GeV$ \cite{Burke:2013yra}. Both temperatures are rescaled due to the fact that the number of degrees of freedom in the non-conformal theory is more than those of 3 favor QCD and one can use $T_{NC}\approx T_{SYM}=3^{-1/3} T_{QCD}$ \cite{Burke:2013yra}.\\
In \tab{Tab:qtab}, the numerical values of the jet quenching parameter is shown for $\mathcal N=4$ SYM theory, non-conformal background ($\phi_M=1$, $\phi_M=3$ and $\phi_M=10$) and experimental data from RHIC and LHC. $\hat{q}$ values are in the units of $GeV^2/fm$. It could be found that the jet quenching results from non-conformal background are in good agreement with the experimental data. It could be found that at $T=0.37 GeV$ varying $\phi_M$ from 1 to 3, the decrease in $\hat{q}$ is $\sim26\%$ while varying $\phi_M$ from 3 to 10, the decrease in $\hat{q}$ is $\sim9\%$. On the other hand, at  $T=0.47 GeV$ varying $\phi_M$ from 1 to 3, the decrease in $\hat{q}$ is $\sim13\%$ while varying $\phi_M$ from 3 to 10, the decrease in $\hat{q}$ is $\sim2\%$. Also, our results imply an increase of the jet-medium interaction at lower temperature where non-conformality effects are magnificent. \\

\begin{table}
\caption{Comparison between the jet quenching values obtained from $\mathcal N=4$ SYM, non-conformal model and experiments. Selected temperature are $T=0.37 GeV$ (from RHIC) and $T=0.47 GeV$ (from LHC).}
\vspace{0.01in}
\centering
\begin{tabular}{|c|c|c|c|c|c|}
\hline
T (GeV) &$\hat{q}_{SYM}$ & $\phi_M=1$ & $\phi_M=3$ & $\phi_M=10$ & Data \\
\hline
$0.37$ & 1.51 & 1.19 & 0.87 & 0.79 & $1.2 \pm 0.3$\\
\hline
$0.47$ & 3.10 & 2.54 & 2.20 & 2.15 & $1.9 \pm 0.7$\\
\hline
\end{tabular}
\label{Tab:qtab}
\end{table}

Holographic models such as \cite{Buchel:2006bv,Gursoy:2010fj} indicate that increasing nonconformality leads to decreasing the value of jet quenching parameter which is in consistency with our results from jet quenching parameter and also with results of light quark stopping distance obtained in the previous section. In general, increasing the deviation from conformal invariance decreases the capability of medium to quench the jets of quarks.  However, in the case of holographic QCD models with phase transition, this parameter exhibits a peak around the QCD phase transition region due to the fact that the system degrees of freedom change rapidly from hadronic gas to QGP \cite{Rougemont:2015wca,Li:2014hja}. 
\section{Summary}
\label{section:Discussion}

The quark-gluon plasma produced in heavy ion collisions is a strongly coupled, non-conformal plasma as indicated by lattice data \cite{Ryu:2015vwa} and hydrodynamics calculations \cite{Bozek:2011if,Schenke:2014zha,Habich:2015rtj,Jeon:2015dfa}. The gauge/gravity duality gives us the opportunity of the study of hot plasma in a strongly coupled regime. 
Suppression of high energy partons produced at heavy ion collision with high transverse momentum is one of the interesting properties of this strongly-coupled plasma that can be investigated using the AdS/CFT correspondence. Although, previous studies show that suppression of jets with sensible energies in $\mathcal N=4$ SYM theory is much larger than its experimental expectation \cite{Morad:2014xla,BitaghsirFadafan:2017tci}. 

In this paper, we investigated the suppression of jet in the strongly-coupled non-conformal plasma in the framework of AdS/CFT correspondence. We considered a holographic five-dimensional model consisting of Einstein gravity coupled to a scalar field with a non-trivial potential which turns out a dual four-dimensional non-conformal gauge theory which exhibits a renormalization group flow between two different fixed points (located at UV and IR) at zero temperature \cite{Attems:2016ugt}. The parameter $\phi_M$ indicated the deviation from conformality as shown in \fig{backgroundsplot}.

We considered a point-like initial condition string created close to the boundary with endpoints that are free to fly apart. The initial conditions of the string are chosen such that the string extends in a direction parallel to the boundary as it falls toward the black hole horizon (falling string). The equations of motion for the string moving the non-conformal background have been solved numerically and plotted in \fig{profile} for a typical string moving in the background with $\phi_M=3$. Since the dynamics of the string depends on the string initial conditions (see \fig{fig:Xtherm}, \fig{fig:StoppingDT} and \fig{fig:StoppingDS} for clarifications), we studied the maximum stopping distance which is insensitive to these IC in the string side. Although, this quantity is not good enough to compute the observables like nuclear modification factor, still is a good quantity to estimate the stopping power of the plasma. 

Our results show that the thermalization distance in non-conformal backgrounds compare to the $\mathcal N=4$ SYM theory depends on the fact that theories have the same temperature, \fig{fig:StoppingDT} or the same entropy, \fig{fig:StoppingDS}. Actually, the \fig{fig:entropy} demonstrates that increasing non-conformality while keeping the entropy fixed leads to hotter plasma and then larger jet suppression. In order to avoid that, we fitted the maximum stopping distance to the well-known relation presented in \eqq{eq:xtherm} and used the $n_{eff}$ as the stopping power of the plasma. Our numerical results, presented in \fig{fig:neff} indicated that the thermalization distance increases by increasing the non-conformality of the plasma. 

In the last section, we calculated the jet quenching parameter for the non-conformal geometry of section \ref{section:nonconBG}. For different values of $\phi_M$, our results are in a good agreement with experimental data. Deviations from conformality are magnificent at lower temperatures and suppress at higher temperatures where non-conformal results approach the corresponding value of the conformal background. As non-conformality increases, the value of jet quenching parameter decreases and this behavior is in consistency with experiments and also with results of light quark stopping distance obtained in section \ref{section:Stoppingdistance}.

\acknowledgments
The authors wish to thank Hesam Soltanpanahi for useful discussions. R. Morad and M. Akbari also wish to thank Prof. M. Maaza and iThemba LABS for the generous support and hospitality.


\begin{thebibliography}{99}
\bibitem{Adams:2005dq} 
  J.~Adams {\it et al.} [STAR Collaboration],
  Nucl.\ Phys.\ A {\bf 757}, 102 (2005)
  doi:10.1016/j.nuclphysa.2005.03.085
  [nucl-ex/0501009].


\bibitem{Adcox:2004mh} 
  K.~Adcox {\it et al.} [PHENIX Collaboration],
  Nucl.\ Phys.\ A {\bf 757}, 184 (2005)
  doi:10.1016/j.nuclphysa.2005.03.086
  [nucl-ex/0410003].


\bibitem{Arsene:2004fa} 
  I.~Arsene {\it et al.} [BRAHMS Collaboration],
  Nucl.\ Phys.\ A {\bf 757}, 1 (2005)
  doi:10.1016/j.nuclphysa.2005.02.130
  [nucl-ex/0410020].


\bibitem{Back:2004je} 
  B.~B.~Back {\it et al.},
  Nucl.\ Phys.\ A {\bf 757}, 28 (2005)
  doi:10.1016/j.nuclphysa.2005.03.084
  [nucl-ex/0410022].


\bibitem{Policastro:2001yc} 
  G.~Policastro, D.~T.~Son and A.~O.~Starinets,
  Phys.\ Rev.\ Lett.\  {\bf 87}, 081601 (2001)
  doi:10.1103/PhysRevLett.87.081601
  [hep-th/0104066].


\bibitem{Kovtun:2004de} 
  P.~Kovtun, D.~T.~Son and A.~O.~Starinets,
  Phys.\ Rev.\ Lett.\  {\bf 94}, 111601 (2005)
  doi:10.1103/PhysRevLett.94.111601
  [hep-th/0405231].


\bibitem{Baier:1996kr} 
  R.~Baier, Y.~L.~Dokshitzer, A.~H.~Mueller, S.~Peigne and D.~Schiff,
  Nucl.\ Phys.\ B {\bf 483}, 291 (1997)
  doi:10.1016/S0550-3213(96)00553-6
  [hep-ph/9607355].


\bibitem{Eskola:2004cr} 
  K.~J.~Eskola, H.~Honkanen, C.~A.~Salgado and U.~A.~Wiedemann,
  Nucl.\ Phys.\ A {\bf 747}, 511 (2005)
  doi:10.1016/j.nuclphysa.2004.09.070
  [hep-ph/0406319].


\bibitem{Maldacena:1997re} 
  J.~M.~Maldacena,
  Int.\ J.\ Theor.\ Phys.\  {\bf 38}, 1113 (1999)
  [Adv.\ Theor.\ Math.\ Phys.\  {\bf 2}, 231 (1998)]
  doi:10.1023/A:1026654312961, 10.4310/ATMP.1998.v2.n2.a1
  [hep-th/9711200].


\bibitem{Witten:1998qj} 
  E.~Witten,
  Adv.\ Theor.\ Math.\ Phys.\  {\bf 2}, 253 (1998)
  doi:10.4310/ATMP.1998.v2.n2.a2
  [hep-th/9802150].


\bibitem{Gubser:1998bc} 
  S.~S.~Gubser, I.~R.~Klebanov and A.~M.~Polyakov,
  Phys.\ Lett.\ B {\bf 428}, 105 (1998)
  doi:10.1016/S0370-2693(98)00377-3
  [hep-th/9802109].


\bibitem{Aharony:1999ti} 
  O.~Aharony, S.~S.~Gubser, J.~M.~Maldacena, H.~Ooguri and Y.~Oz,
  Phys.\ Rept.\  {\bf 323}, 183 (2000)
  doi:10.1016/S0370-1573(99)00083-6
  [hep-th/9905111].


\bibitem{CasalderreySolana:2011us} 
  J.~Casalderrey-Solana, H.~Liu, D.~Mateos, K.~Rajagopal and U.~A.~Wiedemann,
  book:Gauge/String Duality, Hot QCD and Heavy Ion Collisions. Cambridge, UK: Cambridge University Press, 2014
  doi:10.1017/CBO9781139136747
  [arXiv:1101.0618 [hep-th]].


\bibitem{Shuryak:2008eq} 
  E.~Shuryak,
  Prog.\ Part.\ Nucl.\ Phys.\  {\bf 62}, 48 (2009)
  doi:10.1016/j.ppnp.2008.09.001
  [arXiv:0807.3033 [hep-ph]].


\bibitem{Shuryak:2004cy} 
  E.~V.~Shuryak,
  Nucl.\ Phys.\ A {\bf 750}, 64 (2005)
  doi:10.1016/j.nuclphysa.2004.10.022
  [hep-ph/0405066].


\bibitem{Kovtun:2003wp} 
  P.~Kovtun, D.~T.~Son and A.~O.~Starinets,
  JHEP {\bf 0310}, 064 (2003)
  doi:10.1088/1126-6708/2003/10/064
  [hep-th/0309213].


\bibitem{Buchel:2003tz} 
  A.~Buchel and J.~T.~Liu,
  Phys.\ Rev.\ Lett.\  {\bf 93}, 090602 (2004)
  doi:10.1103/PhysRevLett.93.090602
  [hep-th/0311175].


\bibitem{Teaney:2003kp} 
  D.~Teaney,
  Phys.\ Rev.\ C {\bf 68}, 034913 (2003)
  doi:10.1103/PhysRevC.68.034913
  [nucl-th/0301099].


\bibitem{Ackermann:2000tr} 
  K.~H.~Ackermann {\it et al.} [STAR Collaboration],
  Phys.\ Rev.\ Lett.\  {\bf 86}, 402 (2001)
  doi:10.1103/PhysRevLett.86.402
  [nucl-ex/0009011].


\bibitem{Adler:2003kt} 
  S.~S.~Adler {\it et al.} [PHENIX Collaboration],
  Phys.\ Rev.\ Lett.\  {\bf 91}, 182301 (2003)
  doi:10.1103/PhysRevLett.91.182301
  [nucl-ex/0305013].


\bibitem{Back:2004mh} 
  B.~B.~Back {\it et al.} [PHOBOS Collaboration],
  Phys.\ Rev.\ C {\bf 72}, 051901 (2005)
  doi:10.1103/PhysRevC.72.051901
  [nucl-ex/0407012].


\bibitem{ATLAS:2012at} 
  G.~Aad {\it et al.} [ATLAS Collaboration],
  Phys.\ Rev.\ C {\bf 86}, 014907 (2012)
  doi:10.1103/PhysRevC.86.014907
  [arXiv:1203.3087 [hep-ex]].


\bibitem{Chatrchyan:2012ta} 
  S.~Chatrchyan {\it et al.} [CMS Collaboration],
  Phys.\ Rev.\ C {\bf 87}, no. 1, 014902 (2013)
  doi:10.1103/PhysRevC.87.014902
  [arXiv:1204.1409 [nucl-ex]].


\bibitem{Aamodt:2010pa} 
  K.~Aamodt {\it et al.} [ALICE Collaboration],
  Phys.\ Rev.\ Lett.\  {\bf 105}, 252302 (2010)
  doi:10.1103/PhysRevLett.105.252302
  [arXiv:1011.3914 [nucl-ex]].


\bibitem{Adam:2016izf} 
  J.~Adam {\it et al.} [ALICE Collaboration],
  Phys.\ Rev.\ Lett.\  {\bf 116}, no. 13, 132302 (2016)
  doi:10.1103/PhysRevLett.116.132302
  [arXiv:1602.01119 [nucl-ex]].


\bibitem{Aad:2014lta} 
  G.~Aad {\it et al.} [ATLAS Collaboration],
  Phys.\ Rev.\ C {\bf 90}, no. 4, 044906 (2014)
  doi:10.1103/PhysRevC.90.044906
  [arXiv:1409.1792 [hep-ex]].


\bibitem{Khachatryan:2015waa} 
  V.~Khachatryan {\it et al.} [CMS Collaboration],
  Phys.\ Rev.\ Lett.\  {\bf 115}, no. 1, 012301 (2015)
  doi:10.1103/PhysRevLett.115.012301
  [arXiv:1502.05382 [nucl-ex]].


\bibitem{Abelev:2014mda} 
  B.~B.~Abelev {\it et al.} [ALICE Collaboration],
  Phys.\ Rev.\ C {\bf 90}, no. 5, 054901 (2014)
  doi:10.1103/PhysRevC.90.054901
  [arXiv:1406.2474 [nucl-ex]].


\bibitem{Aad:2015gqa} 
  G.~Aad {\it et al.} [ATLAS Collaboration],
  Phys.\ Rev.\ Lett.\  {\bf 116}, no. 17, 172301 (2016)
  doi:10.1103/PhysRevLett.116.172301
  [arXiv:1509.04776 [hep-ex]].


\bibitem{Heller:2011ju} 
  M.~P.~Heller, R.~A.~Janik and P.~Witaszczyk,
  Phys.\ Rev.\ Lett.\  {\bf 108}, 201602 (2012)
  doi:10.1103/PhysRevLett.108.201602
  [arXiv:1103.3452 [hep-th]].


\bibitem{Chesler:2009cy} 
  P.~M.~Chesler and L.~G.~Yaffe,
  Phys.\ Rev.\ D {\bf 82}, 026006 (2010)
  doi:10.1103/PhysRevD.82.026006
  [arXiv:0906.4426 [hep-th]].


\bibitem{Chesler:2015wra} 
  P.~M.~Chesler and L.~G.~Yaffe,
  JHEP {\bf 1510}, 070 (2015)
  doi:10.1007/JHEP10(2015)070
  [arXiv:1501.04644 [hep-th]].


\bibitem{Chesler:2013lia} 
  P.~M.~Chesler and L.~G.~Yaffe,
  JHEP {\bf 1407}, 086 (2014)
  doi:10.1007/JHEP07(2014)086
  [arXiv:1309.1439 [hep-th]].


\bibitem{Casalderrey-Solana:2013sxa} 
  J.~Casalderrey-Solana, M.~P.~Heller, D.~Mateos and W.~van der Schee,
  Phys.\ Rev.\ Lett.\  {\bf 112}, no. 22, 221602 (2014)
  doi:10.1103/PhysRevLett.112.221602
  [arXiv:1312.2956 [hep-th]].


\bibitem{Casalderrey-Solana:2013aba} 
  J.~Casalderrey-Solana, M.~P.~Heller, D.~Mateos and W.~van der Schee,
  Phys.\ Rev.\ Lett.\  {\bf 111}, 181601 (2013)
  doi:10.1103/PhysRevLett.111.181601
  [arXiv:1305.4919 [hep-th]].


\bibitem{Chesler:2016ceu} 
  P.~M.~Chesler,
  JHEP {\bf 1603}, 146 (2016)
  doi:10.1007/JHEP03(2016)146
  [arXiv:1601.01583 [hep-th]].


\bibitem{Chesler:2015bba} 
  P.~M.~Chesler,
  Phys.\ Rev.\ Lett.\  {\bf 115}, no. 24, 241602 (2015)
  doi:10.1103/PhysRevLett.115.241602
  [arXiv:1506.02209 [hep-th]].


\bibitem{Ryu:2015vwa} 
  S.~Ryu, J.-F.~Paquet, C.~Shen, G.~S.~Denicol, B.~Schenke, S.~Jeon and C.~Gale,
  Phys.\ Rev.\ Lett.\  {\bf 115}, no. 13, 132301 (2015)
  doi:10.1103/PhysRevLett.115.132301
  [arXiv:1502.01675 [nucl-th]].


\bibitem{Bozek:2011if} 
  P.~Bozek,
  Phys.\ Rev.\ C {\bf 85}, 014911 (2012)
  doi:10.1103/PhysRevC.85.014911
  [arXiv:1112.0915 [hep-ph]].


\bibitem{Schenke:2014zha} 
  B.~Schenke and R.~Venugopalan,
  Phys.\ Rev.\ Lett.\  {\bf 113}, 102301 (2014)
  doi:10.1103/PhysRevLett.113.102301
  [arXiv:1405.3605 [nucl-th]].


\bibitem{Habich:2015rtj} 
  M.~Habich, G.~A.~Miller, P.~Romatschke and W.~Xiang,
  Eur.\ Phys.\ J.\ C {\bf 76}, no. 7, 408 (2016)
  doi:10.1140/epjc/s10052-016-4237-z
  [arXiv:1512.05354 [nucl-th]].


\bibitem{Jeon:2015dfa} 
  S.~Jeon and U.~Heinz,
  Int.\ J.\ Mod.\ Phys.\ E {\bf 24}, no. 10, 1530010 (2015)
  doi:10.1142/S0218301315300106
  [arXiv:1503.03931 [hep-ph]].


\bibitem{Polchinski:2000uf} 
  J.~Polchinski and M.~J.~Strassler,
  hep-th/0003136.


\bibitem{Karch:2002sh} 
  A.~Karch and E.~Katz,
  JHEP {\bf 0206}, 043 (2002)
  doi:10.1088/1126-6708/2002/06/043
  [hep-th/0205236].


\bibitem{Sakai:2004cn} 
  T.~Sakai and S.~Sugimoto,
  Prog.\ Theor.\ Phys.\  {\bf 113}, 843 (2005)
  doi:10.1143/PTP.113.843
  [hep-th/0412141].


\bibitem{Gursoy:2007cb} 
  U.~Gursoy and E.~Kiritsis,
  JHEP {\bf 0802}, 032 (2008)
  doi:10.1088/1126-6708/2008/02/032
  [arXiv:0707.1324 [hep-th]].


\bibitem{Galow:2009kw} 
  B.~Galow, E.~Megias, J.~Nian and H.~J.~Pirner,
  Nucl.\ Phys.\ B {\bf 834}, 330 (2010)
  doi:10.1016/j.nuclphysb.2010.03.022
  [arXiv:0911.0627 [hep-ph]].


\bibitem{Attems:2016ugt} 
  M.~Attems, J.~Casalderrey-Solana, D.~Mateos, I.~Papadimitriou, D.~Santos-Oliván, C.~F.~Sopuerta, M.~Triana and M.~Zilhão,
  JHEP {\bf 1610}, 155 (2016)
  doi:10.1007/JHEP10(2016)155
  [arXiv:1603.01254 [hep-th]].


\bibitem{Gubser:2008ny} 
  S.~S.~Gubser and A.~Nellore,
  Phys.\ Rev.\ D {\bf 78}, 086007 (2008)
  doi:10.1103/PhysRevD.78.086007
  [arXiv:0804.0434 [hep-th]].


\bibitem{Yin:2013zea} 
  Z.~B.~Yin [ALICE Collaboration],
  Acta Phys.\ Polon.\ Supp.\  {\bf 6}, 479 (2013).
  doi:10.5506/APhysPolBSupp.6.479


\bibitem{Aad:2010bu} 
  G.~Aad {\it et al.} [ATLAS Collaboration],
  Phys.\ Rev.\ Lett.\  {\bf 105}, 252303 (2010)
  doi:10.1103/PhysRevLett.105.252303
  [arXiv:1011.6182 [hep-ex]].


\bibitem{Chatrchyan:2011sx} 
  S.~Chatrchyan {\it et al.} [CMS Collaboration],
  Phys.\ Rev.\ C {\bf 84}, 024906 (2011)
  doi:10.1103/PhysRevC.84.024906
  [arXiv:1102.1957 [nucl-ex]].


\bibitem{Herzog:2006gh} 
  C.~P.~Herzog, A.~Karch, P.~Kovtun, C.~Kozcaz and L.~G.~Yaffe,
  JHEP {\bf 0607}, 013 (2006)
  doi:10.1088/1126-6708/2006/07/013
  [hep-th/0605158].


\bibitem{Gubser:2006bz} 
  S.~S.~Gubser,
  Phys.\ Rev.\ D {\bf 74}, 126005 (2006)
  doi:10.1103/PhysRevD.74.126005
  [hep-th/0605182].


\bibitem{Horowitz:2009pw} 
  W.~A.~Horowitz and Y.~V.~Kovchegov,
  Phys.\ Lett.\ B {\bf 680}, 56 (2009)
  doi:10.1016/j.physletb.2009.07.077
  [arXiv:0904.2536 [hep-th]].


\bibitem{Fadafan:2008bq} 
  K.~Bitaghsir Fadafan, H.~Liu, K.~Rajagopal and U.~A.~Wiedemann,
  Eur.\ Phys.\ J.\ C {\bf 61}, 553 (2009)
  doi:10.1140/epjc/s10052-009-0885-6
  [arXiv:0809.2869 [hep-ph]].


\bibitem{Horowitz:2015dta} 
  W.~A.~Horowitz,
  Phys.\ Rev.\ D {\bf 91}, no. 8, 085019 (2015)
  doi:10.1103/PhysRevD.91.085019
  [arXiv:1501.04693 [hep-ph]].


\bibitem{Chesler:2008uy} 
  P.~M.~Chesler, K.~Jensen, A.~Karch and L.~G.~Yaffe,
  Phys.\ Rev.\ D {\bf 79}, 125015 (2009)
  doi:10.1103/PhysRevD.79.125015
  [arXiv:0810.1985 [hep-th]].


\bibitem{Morad:2014xla} 
  R.~Morad and W.~A.~Horowitz,
  JHEP {\bf 1411}, 017 (2014)
  doi:10.1007/JHEP11(2014)017
  [arXiv:1409.7545 [hep-th]].


\bibitem{Ficnar:2011yj} 
  A.~Ficnar, J.~Noronha and M.~Gyulassy,
  J.\ Phys.\ G {\bf 38}, 124176 (2011)
  doi:10.1088/0954-3899/38/12/124176
  [arXiv:1106.6303 [hep-ph]].


\bibitem{BitaghsirFadafan:2017tci} 
  K.~Bitaghsir Fadafan and R.~Morad,
  Eur.\ Phys.\ J.\ C {\bf 78}, no. 1, 16 (2018)
  doi:10.1140/epjc/s10052-018-5520-y
  [arXiv:1710.06417 [hep-th]].


\bibitem{DEramo:2010wup} 
  F.~D'Eramo, H.~Liu and K.~Rajagopal,
  Phys.\ Rev.\ D {\bf 84}, 065015 (2011)
  doi:10.1103/PhysRevD.84.065015
  [arXiv:1006.1367 [hep-ph]].


\bibitem{Liu:2006he} 
  H.~Liu, K.~Rajagopal and U.~A.~Wiedemann,
  JHEP {\bf 0703}, 066 (2007)
  doi:10.1088/1126-6708/2007/03/066
  [hep-ph/0612168].


\bibitem{Caceres:2006as} 
  E.~Caceres and A.~Guijosa,
  JHEP {\bf 0612}, 068 (2006)
  doi:10.1088/1126-6708/2006/12/068
  [hep-th/0606134].


\bibitem{Buchel:2006bv} 
  A.~Buchel,
  Phys.\ Rev.\ D {\bf 74}, 046006 (2006)
  doi:10.1103/PhysRevD.74.046006
  [hep-th/0605178].


\bibitem{VazquezPoritz:2006ba} 
  J.~F.~Vazquez-Poritz,
  hep-th/0605296.


\bibitem{Nakano:2006js} 
  E.~Nakano, S.~Teraguchi and W.~Y.~Wen,
  Phys.\ Rev.\ D {\bf 75}, 085016 (2007)
  doi:10.1103/PhysRevD.75.085016
  [hep-ph/0608274].


\bibitem{Avramis:2006ip} 
  S.~D.~Avramis and K.~Sfetsos,
  JHEP {\bf 0701}, 065 (2007)
  doi:10.1088/1126-6708/2007/01/065
  [hep-th/0606190].


\bibitem{Gao:2006uf} 
  Y.~h.~Gao, W.~s.~Xu and D.~f.~Zeng,
  hep-th/0611217.


\bibitem{Armesto:2006zv} 
  N.~Armesto, J.~D.~Edelstein and J.~Mas,
  JHEP {\bf 0609}, 039 (2006)
  doi:10.1088/1126-6708/2006/09/039
  [hep-ph/0606245].


\bibitem{Lin:2006au} 
  F.~L.~Lin and T.~Matsuo,
  Phys.\ Lett.\ B {\bf 641}, 45 (2006)
  doi:10.1016/j.physletb.2006.08.024
  [hep-th/0606136].


\bibitem{Sadeghi:2013dga} 
  J.~Sadeghi and S.~Heshmatian,
  Eur.\ Phys.\ J.\ C {\bf 74}, 3032 (2014)
  doi:10.1140/epjc/s10052-014-3032-y
  [arXiv:1308.5991 [hep-th]].


\bibitem{Gursoy:2010fj} 
  U.~Gursoy, E.~Kiritsis, L.~Mazzanti, G.~Michalogiorgakis and F.~Nitti,
  Lect.\ Notes Phys.\  {\bf 828}, 79 (2011)
  doi:10.1007/978-3-642-04864-7-4
  [arXiv:1006.5461 [hep-th]].


\bibitem{Cai:2012eh} 
  R.~G.~Cai, S.~Chakrabortty, S.~He and L.~Li,
  JHEP {\bf 1302}, 068 (2013)
  doi:10.1007/JHEP02(2013)068
  [arXiv:1209.4512 [hep-th]].


\bibitem{Wang:2016noh} 
  L.~Wang and S.~Y.~Wu,
  Eur.\ Phys.\ J.\ C {\bf 76}, no. 11, 587 (2016)
  doi:10.1140/epjc/s10052-016-4421-1
  [arXiv:1609.03665 [hep-th]].


\bibitem{DeWolfe:2009vs} 
  O.~DeWolfe and C.~Rosen,
  JHEP {\bf 0907}, 022 (2009)
  doi:10.1088/1126-6708/2009/07/022
  [arXiv:0903.1458 [hep-th]].


\bibitem{Fadafan:2008uv} 
  K.~Bitaghsir Fadafan,
  Eur.\ Phys.\ J.\ C {\bf 68}, 505 (2010)
  doi:10.1140/epjc/s10052-010-1375-6
  [arXiv:0809.1336 [hep-th]].


\bibitem{Horowitz:2017nbm} 
  W.~A.~Horowitz,
  Nucl.\ Part.\ Phys.\ Proc.\  {\bf 289-290}, 129 (2017).
  doi:10.1016/j.nuclphysbps.2017.05.026


\bibitem{Bianchi:2001kw} 
  M.~Bianchi, D.~Z.~Freedman and K.~Skenderis,
  Nucl.\ Phys.\ B {\bf 631}, 159 (2002)
  doi:10.1016/S0550-3213(02)00179-7
  [hep-th/0112119].


\bibitem{Mas:2007ng} 
  J.~Mas and J.~Tarrio,
  JHEP {\bf 0705}, 036 (2007)
  doi:10.1088/1126-6708/2007/05/036
  [hep-th/0703093 [HEP-TH]].


\bibitem{Chesler:2008wd} 
  P.~M.~Chesler, K.~Jensen and A.~Karch,
  Phys.\ Rev.\ D {\bf 79}, 025021 (2009)
  doi:10.1103/PhysRevD.79.025021
  [arXiv:0804.3110 [hep-th]].


\bibitem{Zakharov:1997uu} 
  B.~G.~Zakharov,
  JETP Lett.\  {\bf 65}, 615 (1997)
  doi:10.1134/1.567389
  [hep-ph/9704255].


\bibitem{Maldacena:1998im} 
  J.~M.~Maldacena,
  Phys.\ Rev.\ Lett.\  {\bf 80}, 4859 (1998)
  doi:10.1103/PhysRevLett.80.4859
  [hep-th/9803002].


\bibitem{Rey:1998ik} 
  S.~J.~Rey and J.~T.~Yee,
  Eur.\ Phys.\ J.\ C {\bf 22}, 379 (2001)
  doi:10.1007/s100520100799
  [hep-th/9803001].


\bibitem{Rey:1998bq} 
  S.~J.~Rey, S.~Theisen and J.~T.~Yee,
  Nucl.\ Phys.\ B {\bf 527}, 171 (1998)
  doi:10.1016/S0550-3213(98)00471-4
  [hep-th/9803135].


\bibitem{Brandhuber:1998bs} 
  A.~Brandhuber, N.~Itzhaki, J.~Sonnenschein and S.~Yankielowicz,
  Phys.\ Lett.\ B {\bf 434}, 36 (1998)
  doi:10.1016/S0370-2693(98)00730-8
  [hep-th/9803137].


\bibitem{Sonnenschein:1999if} 
  J.~Sonnenschein,
  hep-th/0003032.


\bibitem{Liu:2006ug} 
  H.~Liu, K.~Rajagopal and U.~A.~Wiedemann,
  Phys.\ Rev.\ Lett.\  {\bf 97}, 182301 (2006)
  doi:10.1103/PhysRevLett.97.182301
  [hep-ph/0605178].


\bibitem{Burke:2013yra} 
  K.~M.~Burke {\it et al.} [JET Collaboration],
  Phys.\ Rev.\ C {\bf 90}, no. 1, 014909 (2014)
  doi:10.1103/PhysRevC.90.014909
  [arXiv:1312.5003 [nucl-th]].
  
\bibitem{Rougemont:2015wca} 
  R.~Rougemont, A.~Ficnar, S.~Finazzo and J.~Noronha,
  JHEP {\bf 1604}, 102 (2016)
  doi:10.1007/JHEP04(2016)102
  [arXiv:1507.06556 [hep-th]].
  
\bibitem{Li:2014hja} 
  D.~Li, J.~Liao and M.~Huang,
  Phys.\ Rev.\ D {\bf 89}, no. 12, 126006 (2014)
  doi:10.1103/PhysRevD.89.126006
  [arXiv:1401.2035 [hep-ph]].

\end{thebibliography}
\end{document}